\documentclass[traditabstract]{aa}

\usepackage{txfonts}
\usepackage{graphicx}
\usepackage{epsfig,epsf,graphics,lscape,url}
\usepackage{natbib}
\bibpunct{(}{)}{;}{a}{}{,}
\usepackage{longtable}

\begin{document}

\title{Atomic data for neutron-capture elements}
\subtitle{I.  Photoionization and recombination properties of low-charge selenium ions}

\author{N.\ C.\ Sterling$^{\ref{inst1}}$\thanks{NSF Astronomy and Astrophysics Postdoctoral Fellow} \and Michael C.\ Witthoeft$^{\ref{inst2}}$}

\institute{Michigan State University, Department of Physics and Astronomy, 3248 Biomedical Physical Sciences, East Lansing, MI, 48824-2320, USA, \email{sterling@pa.msu.edu} \label{inst1}
\and
NASA Goddard Space Flight Center, Code 662, Greenbelt, MD 20771, USA, \email{michael.c.witthoeft@nasa.gov}\label{inst2}}


\abstract{We present multi-configuration Breit-Pauli AUTOSTRUCTURE calculations of distorted-wave photoionization (PI) cross sections, and total and partial final-state resolved radiative recombination (RR) and dielectronic recombination (DR) rate coefficients for the first six ions of the trans-iron element Se.  These calculations were motivated by the recent detection of Se emission lines in a large number of planetary nebulae.  Se is a potentially useful tracer of neutron-capture nucleosynthesis, but accurate determinations of its abundance in photoionized nebulae have been hindered by the lack of atomic data governing its ionization balance.  Our calculations were carried out in intermediate coupling with semi-relativistic radial wavefunctions.  PI and recombination data were determined for levels within the ground configuration of each ion, and experimental PI cross-section measurements were used to benchmark our results.  For DR, we allowed $\Delta n=0$ core excitations, which are important at photoionized plasma temperatures.  We find that DR is the dominant recombination process for each of these Se ions at temperatures representative of photoionized nebulae ($\sim$10$^4$~K).  In order to estimate the uncertainties of these data, we compared results from three different configuration-interaction expansions for each ion, and also tested the sensitivity of the results to the radial scaling factors in the structure calculations.  We find that the internal uncertainties are typically 30--50\% for the direct PI cross sections and $\sim$10\% for the computed RR rate coefficients, while those for low-temperature DR can be considerably larger (from 15--30\% up to two orders of magnitude) due to the unknown energies of near-threshold autoionization resonances.  These data are available at the CDS, and fitting coefficients to the total RR and DR rate coefficients are presented.  The results are suitable for incorporation into photoionization codes used to numerically simulate astrophysical nebulae, and will enable robust determinations of nebular Se abundances.}


\maketitle

\titlerunning{Atomic Data for Neutron-Capture Elements I}
\authorrunning{N.\ C.\ Sterling \& M.\ C.\ Witthoeft}

\section{Introduction} \label{intro}

The abundances of neutron(\emph{n})-capture elements (atomic number $Z>30$) are sensitive probes of the nucleosynthetic histories of stellar populations and the chemical evolution of the Universe \citep{wally97, busso99, sneden08}.  Our knowledge of the production of these elements is largely based on the interpretation of abundances derived from stellar spectroscopy \citep[e.g.,][and references therein]{smith90, busso99, travaglio04, sneden08, roederer10}.  However, only a limited number of trans-iron elements can be detected in stellar spectra.  Moreover, some classes of stars or stages of evolution with high mass-loss rates are very difficult to study spectroscopically due to photospheric obscuration.

The recent detection of emission lines from \emph{n}-capture elements in a large number of planetary nebulae (PNe) \citep{sterling07, sharpee07, sterling08, sterling09}, as well as other astrophysical objects including  H~II regions \citep[e.g.,][]{aspin94, lumsden96, luhman98, baldwin00, puxley00, okumura01, blum08, roman-lopes09}, starburst galaxies \citep{vanzi08}, and active galactic nuclei \citep{thompson78}, has demonstrated that nebular spectroscopy is a potentially powerful tool for investigating the nucleosynthesis and chemical evolution of trans-iron species.  The abundances of \emph{n}-capture elements are of particular interest in PNe, since these species can be produced in PN progenitor stars via slow \emph{n}-capture nucleosynthesis (the ``\emph{s}-process'') during the asymptotic giant branch (AGB) stage of evolution.  

Nebular spectroscopy provides access to many elements that cannot be detected in cool giant stars, such as the lightest \emph{n}-capture elements ($Z=31$--36) and noble gases.  Due to the difficulty in detecting these species in stellar spectra or supernova remnants \citep{wally95}, the origins of light \emph{n}-capture elements are based predominantly on theoretical considerations that lack empirical validation.  In addition, nebular spectroscopy enables investigations of nucleosynthesis in classes or evolutionary stages of stars obscured by optically thick circumstellar envelopes.  For example, intermediate-mass AGB stars (4--8~M$_{\odot}$) experience heavy mass loss, shielding their photospheres in a cocoon of circumstellar material that hinders spectroscopic investigations in the wavelength regimes of many useful \emph{n}-capture element transitions.  As a result, the contribution of these stars to the Galactic inventory of heavy element nuclei is poorly understood \citep[e.g.,][]{karakas09}.  However, these intermediate-mass stars produce PNe in which \emph{n}-capture element emission lines are readily detected \citep{sharpee07, sterling08}.  Likewise, AGB stars that become carbon-rich as a result of convective dredge-up \citep[see][]{busso99} are characterized by high opacities \citep{marigo02} that substantially increase their mass-loss rates, and hence \emph{s}-process enrichments engendered by the final stages of AGB evolution are not well-constrained.  PNe are composed of material from the stellar envelope at the end of the AGB phase, and hence are useful probes of \emph{s}-process enrichments during late AGB evolution.

However, the accuracy of nebular \emph{n}-capture element abundances is hindered by the poorly known atomic data for these species, thereby limiting the usefulness of nebular spectroscopy for studying the origins of these elements.  The reason for this is that generally only one or two ions of \emph{n}-capture elements have been detected in individual nebulae, and hence the abundances of unobserved ions must be estimated in order to derive elemental abundances.  This is most robustly achieved by numerically simulating the thermal and ionization structure of nebulae with photoionization codes such as Cloudy \citep{ferland98}, but the reliability of these models strongly depends on the availability of accurate atomic data for processes that control the ionization balance.  For photoionized nebulae such as PNe, these data include photoionization (PI) cross sections and rate coefficients for radiative recombination (RR), dielectronic recombination (DR), and charge transfer (CT).  Unfortunately, such data have not been determined for the vast majority of trans-iron element ions.

To address this need, we have computed multi-configuration Breit-Pauli (MCBP) distorted-wave PI cross sections and RR and DR rate coefficients for the first six Se ions, using the atomic structure code AUTOSTRUCTURE \citep{badnell86, badnell97}.  Along with Kr and Xe, Se is one of the most widely observed \emph{n}-capture elements in ionized nebulae \citep{sharpee07, sterling08}.  In fact, Se has been identified in nearly twice as many PNe as any other trans-iron element, and hence is the initial target for our study.  In subsequent papers, we will present similar data for low-charge Kr and Xe ions, as well as CT rate coefficients for several \emph{n}-capture elements.  These data will be suitable for incorporation into photoionization codes such as Cloudy, which can be used to robustly determine correction factors for the abundances of unobserved ions, enabling much more accurate abundance determinations of \emph{n}-capture elements in ionized nebulae than previously possible.

Until our study, the photoionization and recombination properties of Se ions had received very little attention, aside from a handful of PI studies of neutral Se \citep{manson79, gibson86, chen94}, which is a trace species in ionized nebulae such as PNe.  We therefore utilize recently conducted experimental PI cross-section measurements of Se ions \citep{esteves09, sterling11, esteves10, esteves11b} as a comparison and benchmark to our calculations.

We estimate the uncertainties in our computed data by utilizing three different configuration-interaction (CI) expansions for each ion, and test the sensitivity of our results to various internal parameters of the AUTOSTRUCTURE code.  Upon completion of our theoretical study of Se, Kr, and Xe ions we will use Monte Carlo simulations with photoionization codes to reveal the effects of these atomic data uncertainties on nebular abundance determinations, illustrating which systems and atomic processes require further theoretical and/or experimental analysis.

This paper is organized as follows: in Sect.~\ref{calcs}, we provide details of our calculations, and in Sect.~\ref{results} we present our computed PI cross sections and RR and DR rate coefficients along with their associated uncertainties.  Finally, in Sect.~\ref{summ} we summarize our results and provide concluding remarks.

\section{Calculations and methodology}\label{calcs}

We have computed the electronic structure and MCBP distorted-wave PI cross sections for Se$^0$--Se$^{5+}$ with the AUTOSTRUCTURE code \citep{badnell86, badnell97}.  Higher charge states were not considered since they are negligibly populated in PNe, whose central stars are not sufficiently hot \citep[generally $<2\times10^5$~K;][]{napiwot99, stang02} to significantly ionize species with ionization thresholds greater than 100~eV.  RR and DR rate coefficients were determined from the direct and resonant portions of the PI cross sections, respectively, using detailed balance.  We used the independent processes approximation \citep{pindzola92} to treat RR and DR separately.

The theoretical background of such calculations have been provided elsewhere \citep{badnell03, badnell06b}, and we refer the reader to those references for a full discussion.  In each of the following subsections, we discuss details of calculations for each process considered.

\subsection{Electronic structure}

To test the sensitivity of our results to the adopted CI expansions, we constructed three different CI expansions for each Se ion (designated ``small,'' ``medium,'' and ``large'').  The medium configuration sets provide the best compromise between accuracy and computational expense, and are adopted as the basis for our PI, RR, and DR results.  The CI expansions for each ion are listed in Table~\ref{ciexp}.\footnote{Tables~1-4 are available in the online version of this article.}

In all cases, the electronic structure was computed with Thomas-Fermi-Dirac-Amaldi model potentials, adopting intermediate-coupling and $\kappa$-averaged relativistic wavefunctions \citep{cowan76}.  We Schmidt orthogonalized the $\kappa$-averaged orbitals in our calculations, but compared our PI and RR results with those obtained without forcing the radial orbitals to be orthogonal.  The average of the LS term energies were optimized by varying the orbital radial scaling parameters so as to best reproduce experimental energy levels and ionization potentials from NIST \citep{NIST}.  The radial scaling parameters adopted for each CI expansion of each ion are given in Table~\ref{lambdas}.  A comparison of selected calculated and experimental energies is shown in Table~\ref{ecomp}.  We also compare our Einstein A-coefficients with those available in the literature in Table~\ref{acomp}.

The structure computed with the medium configuration sets generally reproduce experimental energies and ionization potentials to a good degree of accuracy, lending credence to our use of orthogonalized radial orbitals.  The ionization potentials are within 2--3\% of the NIST values for all ions with the exception of Se$^0$, which is a very challenging system to model but fortunately is negligibly populated in ionized nebulae such as PNe.  The energies of most levels tabulated by NIST are reproduced to within 5\% for Se$^+$--Se$^{6+}$, and to within 9\% for the neutral case.  A few exceptions occur, particularly for levels within the ground configuration, where the discrepancies with experiment are as large as 23\% for Se$^+$ 4s$^2$\,4p$^3$~$^2$D$_{3/2}$.  


The accuracies of the level energies affects those of the Einstein A-coefficients $A_{ij}$ we computed.  For forbidden transitions within the ground configurations, the main comparisons are the theoretical studies of Bi\'{e}mont \& Hansen in the 1980s (see Table~\ref{acomp} for references), who utilized the Hartree-Fock with relativistic corrections (HFR) and Hartee-plus-statistical-exchange (HXR) methods.  In many cases, our computed $A_{ij}$ agree with those previous studies to within 30--50\%, but exceptions occur, particularly for the lowest-charge states.  However, it is worth noting that our CI expansions generally are larger than those used by Bi\'emont \& Hansen, and hence include configuration mixing not incorporated in their calculations.

\subsection{Photoionization}

MCBP distorted-wave PI cross sections were computed in intermediate coupling for neutral Se and each of the first five ions.  Cross sections were calculated for each level in the ground configurations, as the vast majority of ions in photoionized plasmas reside in the ground configuration.  The CI expansions listed in Table~\ref{ciexp} were used in these calculations, with radial scaling parameters from Table~\ref{lambdas}.  In the cases of Se$^+$, Se$^{2+}$, Se$^{3+}$, and Se$^{5+}$, experimental PI cross sections measured at the Advanced Light Source (ALS) synchrotron radiation facility at Lawrence Berkeley National Laboratory in California were used to test the accuracy of our computations (see Sect.~\ref{pi_exp}).

PI cross sections were computed up to 100~Ryd for each ion.  The length gauge was used at low energies, and AUTOSTRUCTURE automatically switched to velocity gauge at higher energies.  Because of our use of $\kappa$-averaged relativistic orbitals, it was not possible to use acceleration gauge.  The electron energy mesh was interpolated onto the photon energy mesh using a 4-point Lagrange interpolation.  Photoexcitation-autoionization resonances were included for each system, over a more restricted energy range due to the finer energy mesh required for comparison to experimental measurements.  In all cases, we utilized the radial scaling parameters associated with the ($N+1$)-electron ion, as tests showed that using the target scaling parameters led to very poor agreement with experimental PI cross sections.  Scaling parameters for continuum orbitals were taken to be equal to those of the highest principal quantum number bound orbitals with the same orbital angular momentum for $s$, $p$, and $d$ orbitals, and equal to unity otherwise.

\subsection{Radiative recombination}

Total and partial final-state resolved radiative recombination rate coefficients were computed from the direct PI cross sections using detailed balance.  The CI expansions of the ($N+1$)-electron ions were augmented to include one-electron additions to the target's ground and mixing configurations.  Since it was not possible to use acceleration gauge at high energies, the rate coefficients may be inaccurate at the highest temperatures.  However, we have tested that this does not affect the accuracy at photoionized plasma temperatures (see Sect.~\ref{rr_error}).  The rate coefficients were determined over the temperature range (10$^1-10^7$)$z^2$~K, where $z$ is the charge.  We used a fixed maximum principal quantum number of 1000, and a maximum orbital angular momentum of 4 for these calculations.  For $l> 4$, the analytic hydrogenic radial integrals from the recurrence relations of \citet{burgess64} were employed.

\subsection{Dielectronic recombination}

In photoionized nebulae such as PNe, low-temperature DR is the dominant form of recombination for many species.  We therefore computed DR for $\Delta n=0$ core excitations, neglecting $\Delta n\geq 1$ excitations, which are usually negligible at the temperatures of interest ($\sim$10$^4$~K).  However, for DR of Se$^+$ and Se$^{2+}$ (forming Se$^0$ and Se$^+$, respectively), we allowed core excitations into the $5s$ and $5p$ orbitals since the lowest 4$s^2$\,4$p^k$\,5$s$ and 4$s^2$\,4$p^k$\,5$p$ ($k=2$,3) levels lie below the lowest 4$s^2$\,4$p^k$\,4$d$ levels in energy for these ions.  Due to the intensive computational requirements of DR calculations, we ignored the (small) contributions to the Se$^+$ and Se$^{2+}$ DR rate coefficients from excitations into the $5d$, $5f$, and $5g$ orbitals.


In the case of low-temperature DR, it is critical to have accurate energies for low-energy autoionizing states, in order to determine whether they lie just above the ionization threshold and contribute to DR, or just below threshold in which case they do not.  Unfortunately, the energies of these levels have not been spectroscopically determined for Se or other \emph{n}-capture elements, thereby limiting the accuracy of our calculated DR rate coefficients.  To alleviate this source of uncertainty, we used experimental target energies whenever possible \citep{badnell06a}.  When experimental energies are incomplete for a given term, we assumed the theoretical level splitting.  We did not attempt to adjust the energies of high-energy terms or configurations that had not been experimentally measured, as done by \citet{badnell06a}.

We augmented the CI expansions of the structure calculations by including all relevant core excitations for the target ion, allowing for promotions out of the $4s$ and $4p$ subshells.  One-electron additions to all target configurations were added to the ($N+1$)-electron ion CI expansions to account for final states for the Rydberg electron to radiate into the core. 

We determined the DR rate coefficients over the temperature range (10$^1-10^7$)$z^2$~K for each state within the ground configuration.  Captures into Rydberg states up to $n=1000$ and $l=200$ were computed, with $l\geq 9$ treated hydrogenically for all ions except Se$^{5+}$, in which $l\geq 11$ states were treated hydrogenically to allow the calculation to converge.

\section{Results}\label{results}

\subsection{Photoionization cross sections}

The photoionization and photoexcitation cross sections from AUTOSTRUCTURE were processed with the ADASPI and ADASPE codes, which respectively produce the \textit{adf39} and \textit{adf38} output files \citep[see][for a detailed description]{summers05}.  To simplify the presentation of these results, we used the \textit{xpeppi} post-processing code\footnote{see http://amdpp.phys.strath.ac.uk/autos/} to add the direct PI and photoexcitation-autoionization cross sections.

\begin{figure}
  \resizebox{\hsize}{!}{\includegraphics{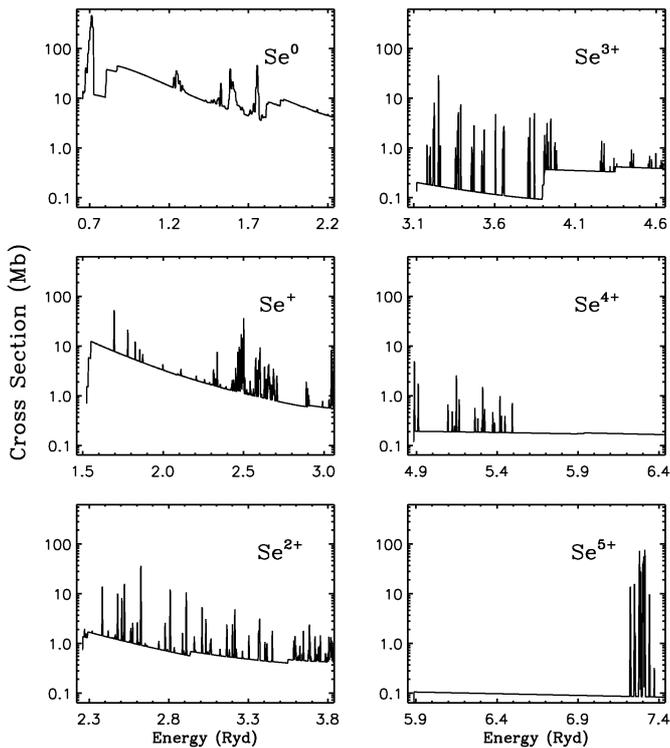}}
  \caption{Ground state PI cross sections for Se$^0$--Se$^{5+}$, in the energy region near the ground state ionization threshold.} \label{allpi}
\end{figure}

PI cross sections from 0 to 100 Rydbergs for Se$^0$--Se$^{5+}$ are presented as supplementary data files available at the CDS, and are illustrated in Fig.~\ref{allpi} for PI out of the ground state of each ion.  The photoionization plus photoexcitation-autoionization cross sections (files labeled ``se$q$+\_XPITOT.dat,'' $q=0-5$) are provided for all states in the ground configuration of each ion.  In those files, column~1 gives the photon energy relative to the ground state and column~2 the photon energy relative to the initial state.  Columns~3 and 4 present the photo-electron energy relative to the ground and initial states, respectively, and in column~5 the cross section is given in Mb.  All energies are in Rydbergs.  The data for different states are delimited by comment lines (denoted by a \# sign in the first column), followed by three numbers indicating the energy of that level relative to the ground state (or the total energy, in the case of the ground state), the total energy of the target state, and the energy order of the initial state (1 for the ground state, 2 for the first excited state, etc.).

We also provide direct PI cross sections without photoexcitation-autoionization resonances, in files labeled ``se$q$+\_XDPITOT.dat.''  In these files, there are two columns, the first of which is the photon energy relative to the ground state, the second the cross section in Mb.  Cross sections for different states in the ground configuration are delimited in the same manner as for the se$q$+\_XPITOT.dat files.

\subsubsection{Estimation of uncertainties}\label{pi_error}

Special effort was made to estimate uncertainties in the PI cross sections, as these uncertainties affect the accuracy of abundances derived from spectra of astrophysical nebulae \citep[e.g., ][]{sterling07}.  We compared the cross sections computed with the three CI expansions for each ion, and also varied internal parameters or tested assumptions in our calculations (e.g., the values of the scaling parameters for continuum orbitals, the sensitivity to the gauge used, and our orthogonalization of the $\kappa$-averaged orbitals).  We also compared our results to experimental measurements, as described in the following subsection.

Because we used $\kappa$-averaged relativistic wavefunctions, it was necessary to compute the cross sections in the velocity gauge even at the highest energies, where acceleration gauge is optimal.  We compared cross sections computed in the length gauge at all energies to those computed solely in velocity gauge.  We found negligible differences ($<<1$\%) in the cross sections at both low and high energies, indicating that our use of velocity gauge up to 100~Ryd is not a significant source of error.

For continuum orbitals, we used scaling parameters equal to those of the highest principal quantum number with the same orbital angular momentum for $s$, $p$, and $d$ orbitals, and unity for higher angular momentum states.  We tested this assumption by calculating PI cross sections with all continuum orbital scaling parameters set to unity.  The direct cross sections from these two calculations agree to within 10\% (often better) near the ground state threshold for all ions considered, although the resonance strengths vary (especially for Se$^+$, where the resonances are weaker by nearly a factor of two when the continuum scaling parameters are 1.0).

For each ion, we also calculated the cross sections without orthogonalizing the radial orbitals.  Compared to our calculations with orthogonalized orbitals, the direct PI cross sections agree to within 15\% near the ground state threshold of Se$^{4+}$, and to better than 10\% for other ions.  However, threshold and resonance energies, as well as resonance strengths, are altered when orthogonalization is not forced, most significantly for the neutral and near-neutral cases.  Threshold energies tend to agree best with NIST values for our calculations with orthogonalized radial orbitals, particularly for the neutral and single ion, in which the thresholds are 0.005--0.02~Ryd higher than in the non-orthogonal calculations.

\begin{figure}
  \resizebox{\hsize}{!}{\includegraphics{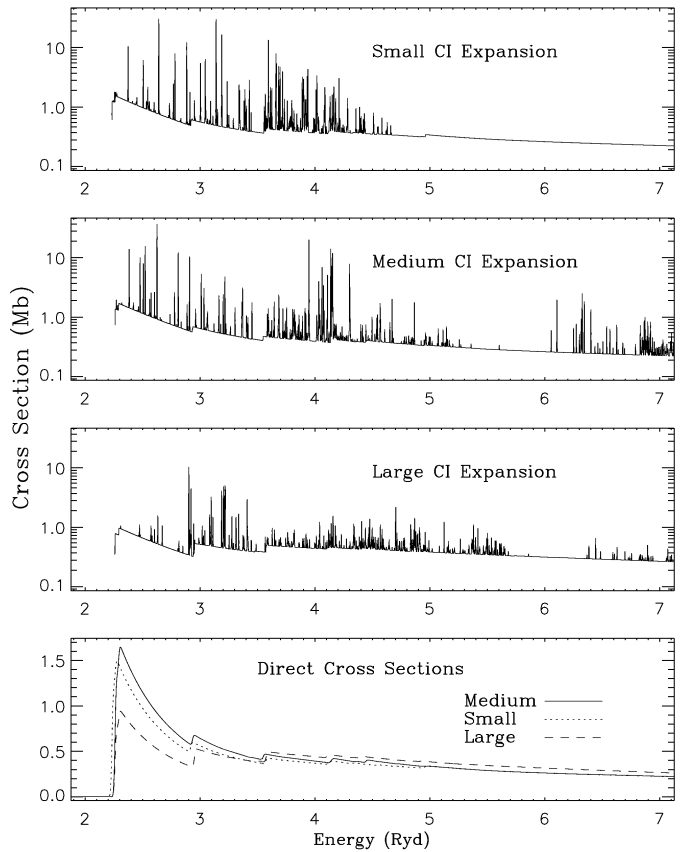}}
  \caption{Comparison of Se$^{2+}$ ground state PI cross sections up to 5~Ryd above the threshold using different CI expansions, as indicated in each of the top three panels.  The bottom panel compares the direct PI cross sections, with photoexcitation-autoionization resonances removed, for the three different CI expansions.} \label{se2comp}
\end{figure}

\begin{figure}
  \resizebox{\hsize}{!}{\includegraphics{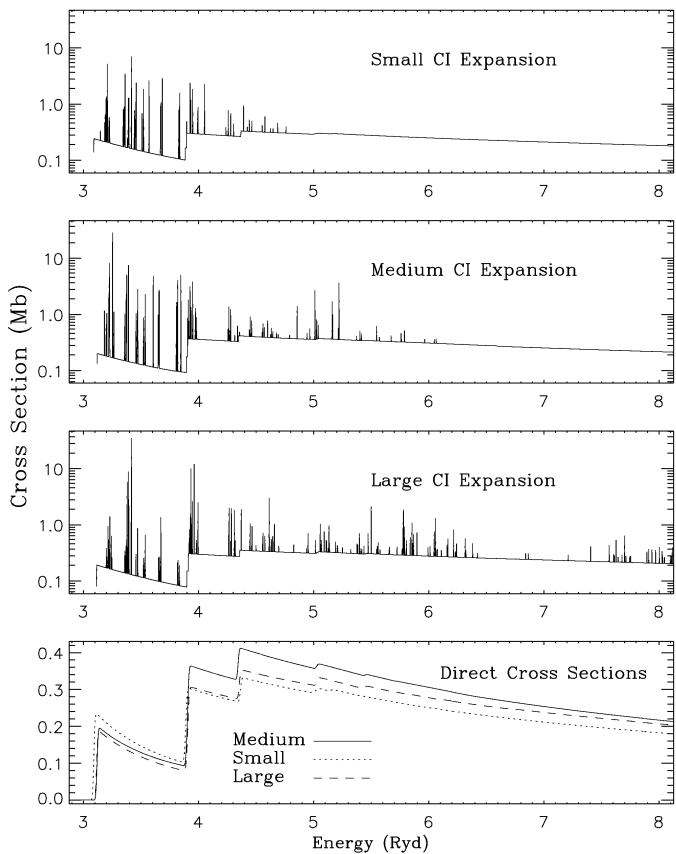}}
  \caption{Comparison of Se$^{3+}$ ground state PI cross sections up to 5~Ryd above the threshold using different CI expansions, as indicated in each of the top three panels.  The bottom panel compares the direct PI cross sections, with photoexcitation-autoionization resonances removed, for the three different CI expansions.} \label{se3comp}
\end{figure}

The most significant differences are found from comparing the PI cross sections calculated with different CI expansions.  The deviations in the electronic structure for the target and ($N+1$)-electron ions lead to disparate resonance energies and strengths, as can be seen in Figs.\ \ref{se2comp} and \ref{se3comp} for the cases of Se$^{2+}$ and Se$^{3+}$.  The sensitivity of the resonance strengths and positions to the CI expansion and other internal parameters illustrates the challenge of accurately reproducing photoexcitation-autoionization resonances in the distorted-wave approximation.

However, given the uncertainty in PI resonance positions from any type of theoretical calculation, numerical models of ionized astrophysical nebulae often only consider direct PI \citep[e.g.,][]{verner96b} or convolve the cross section with Gaussians to smooth over the uncertainty in resonance positions \citep{bautista01}.  Indeed, our goal in this work is to best reproduce the direct PI cross sections for low-charge Se ions, and to quantify their uncertainties within the distorted wave approximation.  

The bottom panels of Figs.\ \ref{se2comp} and \ref{se3comp} compare the direct ground state PI cross sections calculated with each of the three CI expansions.  In general, the cross sections agree to within roughly 30\% for most ions near the ground state ionization threshold.  The discrepancy can be as large as 60\% in the cases of Se$^+$ and Se$^{2+}$, but generally decreases with energy.  For Se$^{4+}$, the direct cross sections agree to within 10\% near threshold.

Given these internal tests of our cross sections and our comparisons to experimental measurements (Sec.\ \ref{pi_exp}), we estimate our calculated direct PI cross sections to be uncertain by 30--50\% in the energy region near the ground state ionization threshold.


\subsubsection{Comparison to experiment} \label{pi_exp}

To benchmark and test the accuracy of our results, we have compared our (medium CI expansion) calculations to experimental absolute PI cross-section measurements performed at the ALS synchrotron radiation facility.  Se$^+$, Se$^{2+}$, Se$^{3+}$, and Se$^{5+}$ cross sections\footnote{The Se$^{4+}$ cross section could not be measured, due to the presence of strong autoionizing states in the parent Se$^{4+}$ ion beam that led to an extremely high background.} were measured from the threshold of the highest-energy metastable state in the ground configuration up to 10~eV beyond the ground state ionization threshold \citep{esteves10}.  Experimental results for Se$^+$ \citep{sterling11} and Se$^{3+}$ \citep{esteves09} have been published, and we compare our theoretical results to experiment for those ions in this section (a similar comparison to the as-yet unpublished Se$^{2+}$ and Se$^{5+}$ cross sections was also performed).

The experimental measurements were described by \citet{esteves09} and \citet{sterling11}, and we refer the reader to those papers for full details.  Briefly, the PI cross sections were measured via the merged beam method \citep{lyon86}.  The photoionization spectra were measured by scanning across photon energies, and normalized to an absolute scale with absolute cross-section measurements performed at discrete energies.  Overall, uncertainties in the experimental cross sections were estimated to be $\sim$30\% for Se$^+$ and $\sim$20\% for Se$^{3+}$ \citep{sterling11, esteves09}.

The experimental data are complicated by the fact that the primary ion beams were composed of an unknown admixture of ground and metastable states.  We therefore linearly combined the computed cross sections of the ground configuration states to best reproduce the direct experimental cross sections and (when possible) families of resonances.

\begin{figure}
  \resizebox{\hsize}{!}{\includegraphics{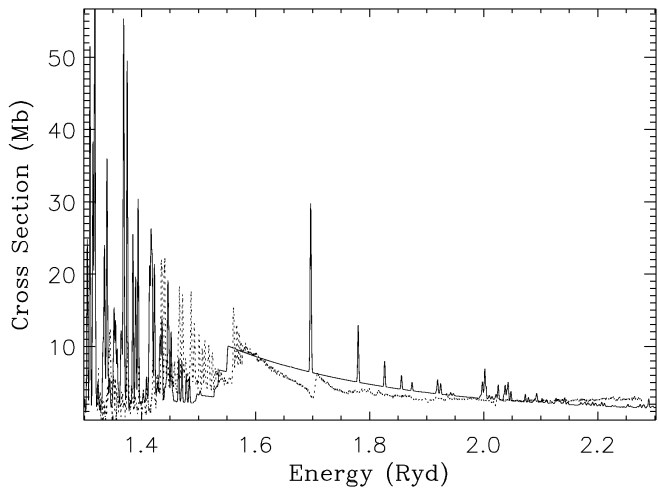}}
  \caption{Comparison of calculated (solid line) and experimental (dotted line) Se$^{+}$ PI cross sections.  The theoretical curve is a linear combination of the cross sections from the ground configuration ($^4$S$_{3/2}$, $^2$D$_{3/2}$, $^2$D$_{5/2}$, $^2$P$_{1/2}$, $^2$P$_{3/2}$) weighted by the factors (0.53, 0.15, 0.05, 0.11, 0.16), respectively.  The theoretical cross section was convolved with a Gaussian of 28~meV FWHM to reproduce the experimental resolution.} \label{se1exp}
\end{figure}
\begin{figure}
  \resizebox{\hsize}{!}{\includegraphics{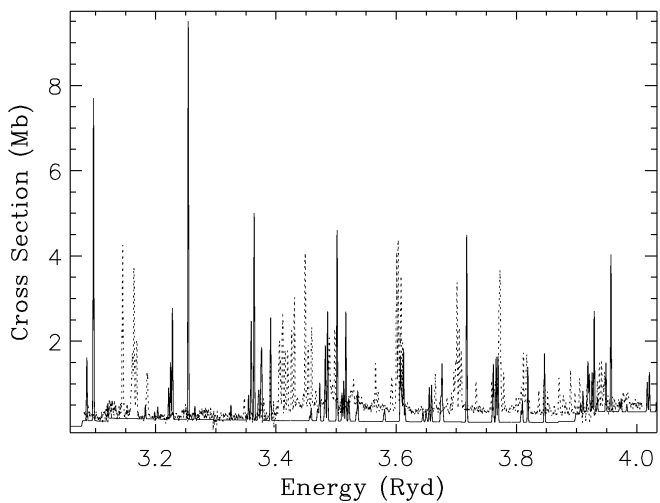}}
  \caption{Comparison of calculated (solid line) and experimental (dotted line) Se$^{3+}$ PI cross sections.  The theoretical curve is a statistically-weighted linear combination of the cross sections from the two ground configuration states.  The theoretical cross section was convolved with a Gaussian of 20~meV FWHM to reproduce the experimental resolution.} \label{se3exp}
\end{figure}

Figs.\ \ref{se1exp} and \ref{se3exp} compare our calculated Se$^+$ and Se$^{3+}$ PI cross sections (solid lines) with experimental measurements (dotted lines).  The computed Se$^+$ cross section was convolved with a Gaussian of 28~meV~FWHM to reproduce the experimental resolution.  By visual comparison to the experimental data, we found best agreement by weighting the contribution of the ground configuration states ($^4$S$_{3/2}$, $^2$D$_{3/2}$, $^2$D$_{5/2}$, $^2$P$_{1/2}$, $^2$P$_{3/2}$) by (0.53, 0.15, 0.05, 0.11, 0.16).  As expected (see Sect.~\ref{pi_error}), our distorted-wave calculations do not accurately reproduce the resonance structure of the experimental PI cross section.  In addition, the Fano profile near 1.7~Ryd is not reproduced by our calculations ($R$~matrix techniques are necessary to reproduce resonance interference effects), and the lack of this interference causes the discrepancy in the direct cross section in this energy region.  The calculated and experimental Se$^+$ direct cross sections agree to within 30--50\% over the range of experimental energies, except above 2.2~Ryd, where the experimental measurements are more uncertain since only a single absolute measurement was made above 1.98~Ryd (compared to absolute measurements at lower energies, where independent estimates were performed 2--3 times for each energy).


The calculated Se$^{3+}$ cross section was convolved with a 20~meV~FWHM Gaussian, and we found that statistical weighting of the two ground configuration states provided reasonable agreement with the experimental measurements of \citet{esteves09} near threshold.  Again, resonance energies and strengths are in poor agreement, though several general resonance families are seen in both the theoretical and experimental results.  The computed direct cross section is smaller than that of experiment, but agrees within the noise up to 3.38~Ryd.  From 3.38 to 3.9~Ryd, however, an apparent threshold is observed in the experimental data that is not reproduced in our theoretical cross section.  In the calculated cross section, the threshold near 3.9~Ryd is due to photoionization into the target Se$^{4+}$~4$s$\,4$p$~$^{3}$P term.  The energy of this term is lower than the NIST value, and hence an inaccurate energy for this threshold cannot explain the experimental threshold at 3.38~Ryd.  Given that 4$s$\,4$p$~$^{3}$P is the lowest-energy excited term of Se$^{4+}$ and that it is unlikely the primary Se$^{3+}$ ion beam contained metastable states with excitation energies of more than one Rydberg, this experimental feature is puzzling.  However, recent revisions to the analysis of the experimental Se$^{3+}$ data indicate that the apparent threshold at 3.38~Ryd is due to a cluster of poorly-resolved resonances, as well as an error in the normalization of the data to an absolute scale at energies above 3.38~Ryd (D.\ A.\ Esteves, private communication).

Comparisons to the as-yet unpublished Se$^{2+}$ and Se$^{5+}$ cross sections (not shown in this paper) are more encouraging, at least in terms of the direct cross sections.  In both cases, our calculations agree within the noise of the experimental data.

\subsection{Radiative recombination rate coefficients}

RR rate coefficients were calculated from the direct photoionization cross sections using the ADASRR code.  The resulting \textit{adf48} output files (available at the CDS) are similar in structure to \textit{adf09} files for DR \citep[see][for a detailed description]{summers05}.  In these files, partial RR rate coefficients to states with $n \leq 8$ are fully $J$-resolved, and are bundled over the final outer quantum numbers as bundled-$nl$ for $n \leq 10$, and bundled-$n$ for $n < 1000$ \citep[see][]{badnell06b}.  At the end of each \textit{adf48} file, total RR rate coefficients are written as a function of temperature for each target ground configuration state.

We fit the rate coefficients with the analytical form \citep{verner96a, badnell06b}
\begin{equation}
\alpha_{\mathrm{RR}}(T) = A \times \left[ \sqrt{T/T_0}(1 + \sqrt{T/T_0})^{1-B_0}(1 + \sqrt{T/T_1})^{1+B_0} \right]^{-1}, \label{rr1}
\end{equation}
where 
\begin{equation}
B_0 = B + C\mathrm{exp}(-T_2/T). \label{rr2}
\end{equation}
In this equation, $T_{0,1,2}$ are in temperature units, $A$ and the rate coefficient $\alpha_{\mathrm{RR}}(T)$ have units of cm$^3$\,s$^{-1}$, and $B$ and $C$ are dimensionless parameters.  For several low-charge states, the RR rate coefficients exhibit large ``bumps'' at high temperature.  These features are caused by electron capture into low principal quantum number states, in which the nucleus is not as effectively screened as for higher states.  Equations~\ref{rr1} and \ref{rr2} are not able to reproduce the size of the bumps, and become inaccurate at high temperatures.  For these Se$^+$--Se$^{3+}$ target states, we therefore fit the high-temperature RR rate coefficients with the function
\begin{equation}
\alpha_{\mathrm{RR}}(T) = \frac{1}{T^{3/2}} \sum_{i=1}^{n} c_i\mathrm{exp}(-E_i/T), \label{dreq}
\end{equation}
where $T$ and $E_i$ are in temperature units (K), the rate coefficient is in cm$^3$\,s$^{-1}$, $c_i$ in cm$^3$\,s$^{-1}$\,K$^{3/2}$, and $n$ ranges from 5 to 7 depending on the ion.  This analytical function is typically used to fit DR rate coefficients \citep[e.g., ][]{zatsarinny03}, as done in Sect.~\ref{dr}.

The fit coefficients were determined using a non-linear least-squares fit algorithm, generally leading to fit accuracies within 5\% for the low-temperature fits, and $\leq$2\% at high temperatures.  However, the low-temperature fits are only accurate to within 7.5\% for Se$^+$ (level~1) and Se$^{3+}$ (level~1), and the accuracies of the high-temperature fits for Se$^+$ are 3\% (levels~2-4) to 4\% (level~5).  Moreover, Equations~\ref{rr1} and \ref{rr2} are unable to reproduce the rate coefficient for RR onto level~5 of Se$^+$ at the lowest temperatures ($\leq 20$~K), where they deviate from the calculated rate coefficients by up to 20\%.  Aside from this exception, the fits have the correct asymptotic forms outside of the temperature range (10$^1-10^7$)$z^2$~K (specifically, Equations~\ref{rr1} and \ref{rr2} describe the asymptotic behavior as $T\rightarrow 0$~K, and Equation~\ref{dreq} correctly describes the rate coefficient at the high-temperature limit).

\begin{figure}
  \resizebox{\hsize}{!}{\includegraphics{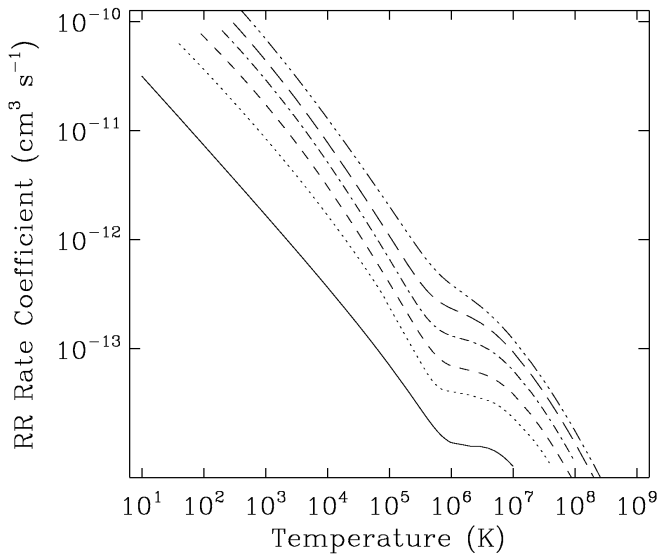}}
  \caption{Radiative recombination rate coefficients as a function of temperature for ground state target ions Se$^+$ (solid curve), Se$^{2+}$ (dotted curve), Se$^{3+}$ (short-dashed curve), Se$^{4+}$ (dash-dot-dash curve), Se$^{5+}$ (long-dashed curve), and Se$^{6+}$ (dash-dot-dot-dot-dash curve).} \label{rrfig}
\end{figure}

Table~\ref{rrlowtfits} displays the fit coefficients from equations~\ref{rr1} and \ref{rr2}, as well as the maximum temperature $T_{\rm max}$ for which the fits are valid to within the stated accuracies.  If $T_{\rm max}$ is not given in Table~\ref{rrlowtfits}, then equations~\ref{rr1} and \ref{rr2} are accurate over the full temperature range (10$^1-10^7$)$z^2$~K and correctly describe the asymptotic behavior of the rate coefficients at the high-temperature limit.  Table~\ref{rrhitfits} provides fit coefficients for the RR rate coefficients from $T_{\rm max}$ to $10^7 z^2$~K.  In Fig.\ \ref{rrfig}, we display the ground state RR rate coefficients for low-charge Se ions as a function of temperature.

\addtocounter{table}{4}

\begin{table*}
\centering
\caption{Fit coefficients for radiative recombination rate coefficients at low temperature (see Equations~\ref{rr1} and \ref{rr2}). $T_{\rm max}$ represents the maximum temperature for which these fits are valid to within the stated accuracies (5\%, unless noted).  For recombination at temperatures exceeding $T_{\rm max}$, the fits from Equation~\ref{dreq} and coefficients listed in Table~\ref{rrhitfits} should be used.  If no $T_{\rm max}$ is given, then the fits are accurate over the entire temperature range (10$^1-10^7$)$z^2$~K.  The notation $x(y)$ denotes $x\times 10^y$.}\label{rrlowtfits} 
\begin{tabular}{lcccccccc}
\hline \hline
Target & & $A$ & $B$ & $T_0$ & $T_1$ & $C$ & $T_2$ & $T_{\rm max}$ \\
Ion\tablefootmark{a} & Level & (cm$^3$\,s$^{-1}$) &  & (K) & (K) &  & (K) & (K) \\
\hline
Se$^+$ & 1\tablefootmark{b} & 7.395(-10) & 0.7288 & 7.273(-2) & 3.495(6) & 0.3306 & 2.160(6) & ... \\
 & 2 & 2.808(-11) & 0.4980 & 3.761 & 3.415(129) & 0.3365 & 4.621(1) & 2.0(3) \\
 & 3 & 2.747(-11) & 0.4818 & 3.926 & 2.747(111) & 0.3515 & 4.451(1) & 2.0(3) \\
 & 4\tablefootmark{c} & 1.439(-11) & 0.0535 & 2.036(1) & 6.815(130) & 0.7152 & 4.065(1) & 2.0(3) \\
 & 5\tablefootmark{d} & 3.719(-11) & 0.4765 & 3.786 & 3.019(101) & 0.2935 & 6.541(1) & 1.0(3) \\
Se$^{2+}$ & 1 & 6.253(-11) & 0.6744 & 6.081(1) & 5.440(5) & 0.8633 & 2.232(6) & 8.0(4) \\
 & 2 & 3.804(-12) & 0.5422 & 7.174(3) & 3.930(6) & 1.3093 & 1.915(5) & 8.0(4) \\
 & 3 & 2.977(-11) & 1.0695 & 3.398(1) & 1.907(5) & 0.9534 & 2.179(6) & 8.0(4) \\
 & 4 & 1.358(-11) & 1.1128 & 7.719(1) & 2.460(5) & 1.0094 & 2.152(6) & 8.0(4) \\
 & 5 & 2.109(-11) & 1.0811 & 2.623(1) & 2.879(5) & 0.8678 & 2.142(6) & 8.0(4) \\
Se$^{3+}$ & 1\tablefootmark{e} & 1.496(-11) & 0.0000 & 3.270(3) & 9.004(7) & 0.7282 & 1.394(6) & ... \\
 & 2 & 5.353(-11) & 0.9825 & 5.028(1) & 6.721(5) & 1.0127 & 4.915(5) & 9.0(5) \\
Se$^{4+}$ & 1 & 2.664(-11) & 0.0000 & 3.018(3) & 1.142(8) & 0.6892 & 1.136(6) & ... \\
Se$^{5+}$ & 1 & 5.822(-11) & 0.1471 & 1.562(3) & 1.905(8) & 0.5119 & 1.031(6) & ... \\
Se$^{6+}$ & 1 & 1.293(-10) & 0.3096 & 8.107(2) & 2.554(8) & 0.3114 & 1.278(6) & ... \\
\hline
\end{tabular}
\tablefoot{
\tablefoottext{a}{Note that the ion and level numbers correspond to the target ion (i.e., before recombination).}
\tablefoottext{b}{The fit is accurate to within 7.5\%.}
\tablefoottext{c}{The fit is accurate to within 5\% except at the lowest temperature (10~K), where it is accurate to within 7\%.}
\tablefoottext{d}{The fit is accurate to within 5\% except at the lowest temperatures (below 20~K), where the fit agreement can be as poor as 20\%.  For this reason, the fit to the RR rate coefficent for this state should not be considered valid below 20~K.}
\tablefoottext{e}{The fit is accurate to within 7\%.}
}
\end{table*}

\begin{table*}
\centering
\caption{Fit coefficients for radiative recombination rate coefficients at high temperatures, valid from $T_{\rm max}$ (see Table~\ref{rrlowtfits}) to $10^7z^2$~K.  See Equation~\ref{dreq}.  Coefficients $c_{\rm i}$ are in cm$^3$\,s$^{-1}$\,K$^{3/2}$, and $E_{\rm i}$ are in K.  States whose RR rate coefficients are accurately described by Equations~\ref{rr1} and \ref{rr2} (with coefficients given in Table~\ref{rrlowtfits}) over the full temperature range (10$^1-10^7$)$z^2$~K are not listed here.  The fits are accurate to within 3\% for Se$^+$ levels~2--4, 4\% for Se$^+$ level~5, and $\leq$2\% otherwise.  The notation $x(y)$ denotes $x\times 10^y$.}\label{rrhitfits} 
\begin{tabular}{lcccccccc}
\hline \hline
Target & &  &  &  &  &  &  &  \\
Ion\tablefootmark{a} & Level & $c_1$ & $c_2$ & $c_3$ & $c_4$ & $c_5$ & $c_6$ & $c_7$ \\
\hline
Se$^+$ & 2 & 6.380(-8) & 3.641(-7) & 1.412(-6) & 2.856(-6) & 6.461(-6) & 1.067(-4) & 4.751(-4) \\
 & 3 & 6.425(-8) & 3.638(-7) & 1.403(-6) & 2.811(-6) & 6.138(-6) & 1.018(-4) & 4.729(-4) \\
 & 4 & 7.502(-8) & 4.203(-7) & 1.429(-6) & 2.582(-6) & 5.538(-6) & 8.929(-5) & 3.948(-4) \\
 & 5 & 3.371(-8) & 1.857(-7) & 8.251(-7) & 2.939(-6) & 5.302(-6) & 8.719(-5) & 3.936(-4) \\
Se$^{2+}$ & 1 & 2.061(-6) & 8.034(-6) & 1.447(-5) & 2.769(-4) & 1.237(-3) & 2.609(-3) & ... \\
 & 2 & 2.232(-6) & 8.551(-6) & 1.552(-5) & 2.780(-4) & 1.240(-3) & 2.586(-3) & ... \\
 & 3 & 1.018(-5) & 1.613(-5) & 2.749(-4) & 1.236(-3) & 2.612(-3) & ... & ... \\
 & 4 & 1.195(-5) & 7.588(-5) & 6.316(-4) & 1.965(-3) & 4.432(-3) & ... & ... \\
 & 5 & 5.306(-6) & 7.409(-6) & 7.781(-5) & 5.502(-4) & 1.469(-3) & 2.035(-3) & ... \\
Se$^{3+}$ & 2 & 4.289(-5) & 1.638(-3) & 1.161(-2) & 5.245(-3) & 2.772(-2) & ... & ... \\
\hline
 & & $E_1$ & $E_2$ & $E_3$ & $E_4$ & $E_5$ & $E_6$ & $E_7$ \\
\hline
Se$^+$ & 2 & 6.920(2) & 6.054(3) & 2.936(4) & 1.100(5) & 6.510(5) & 3.218(6) & 1.207(7) \\
 & 3 & 6.990(2) & 6.080(3) & 2.926(4) & 1.087(5) & 6.306(5) & 3.137(6) & 1.190(7) \\
 & 4 & 6.984(2) & 6.165(3) & 2.893(4) & 1.029(5) & 6.244(5) & 3.151(6) & 1.194(7) \\
 & 5 & 3.045(2) & 2.838(3) & 1.405(4) & 6.763(4) & 5.084(5) & 3.075(6) & 1.182(7) \\
Se$^{2+}$ & 1 & 3.108(3) & 4.135(4) & 3.780(5) & 2.581(6) & 1.008(7) & 3.998(7) & ... \\
 & 2 & 4.231(3) & 4.997(4) & 4.748(5) & 2.637(6) & 1.015(7) & 4.037(7) & ... \\
 & 3 & 3.628(4) & 4.849(5) & 2.644(6) & 1.019(7) & 4.124(7) & ... & ... \\
 & 4 & 6.274(4) & 1.334(6) & 5.185(6) & 2.057(7) & 1.135(8) & ... & ... \\
 & 5 & 2.145(4) & 1.718(5) & 1.457(6) & 5.086(6) & 1.786(7) & 6.142(7) & ... \\
Se$^{3+}$ & 2 & 6.867(4) & 1.119(6) & 3.421(6) & 2.101(7) & 9.183(7) & ... & ... \\
\hline
\end{tabular}
\tablefoot{
\tablefoottext{a}{Note that the ion and level numbers correspond to the target ion (i.e., before recombination).}
}
\end{table*}

\subsubsection{Estimation of uncertainties} \label{rr_error}

We tested the sensitivity of the calculated RR rate coefficients to the use of different CI expansions and other internal parameters in our calculations.  The uncertainties that we discuss were estimated at temperatures near 10$^4$~K, similar to electron temperatures in photoionized nebulae.

As is the case for PI, the RR rate coefficients are most sensitive to the CI expansion used.  The rate coefficients calculated with the small and large configuration sets (Table~\ref{ciexp}) differ from the presented rate coefficients (from the medium CI expansion) typically by less than 10\% near 10$^4$~K.  The sole exception is that of Se$^+$, for which the deviation from the presented rate coefficients is 25--30\%.

Other tested parameters produced smaller changes in the RR rate coefficients.  For example, to test the sensitivity of forcing velocity gauge at high energies, we computed the rate coefficients from PI cross sections cut off at the maximum velocity gauge energy (beyond which velocity gauge must be forced in AUTOSTRUCTURE).  The RR rate coefficients were unaffected except at the highest temperatures (typically a few times 10$^5$--10$^6$~K).  This indicates that uncertainties associated to our use of velocity gauge at high energies are negligible for the computed RR rate coefficients at photoionized plasma temperatures, and become important only above 10$^5$~K for singly- and doubly-charged Se, and above 10$^6$~K for higher charge states.

The calculations were performed using 3 interpolation energies per decade.  We tested the effects of using a finer interpolation mesh of five energies per decade, and found differences of less than 1\% in the RR rate coefficients for all of the investigated Se ions.  Similarly, the use of target ion scaling parameters versus those of the $N+1$-electron ion produced discrepancies in the rate coefficients larger than 1\% only in the case of Se$^+$ and Se$^{2+}$, where the differences are 3--7\% (depending on the state) near 10$^4$~K.  Finally, we computed RR rate coefficients without Schmidt orthogonalizing the radial orbitals, and found that the rate coefficient differed by a maximum of 1--3\% (for Se$^+$ and Se$^{2+}$) compared to our orthogonalized calculations.

We conclude that our computed RR rate coefficients have internal uncertainties of less than 10\% with the exception of Se$^+$, where the uncertainty is 30--40\%.

\subsection{Dielectronic recombination rate coefficients}\label{dr}

DR rate coefficients were computed from the resonant photoionization cross sections with the ADASDR code, and the \textit{adf09} output files are available at the CDS \citep[see][for a detailed description of the structure of these files]{summers05}.  The rate coefficients were fitted with functions of the form Equation~\ref{dreq}, with $n$ ranging between 5 and 7 (see Table~\ref{drfits}).  The fits for most ions are accurate to within 5\% over the temperature range (10$^1-10^7$)$z^2$~K, and reproduce the correct asymptotic behavior outside of this temperature range.  The fitting algorithm was not able to accurately reproduce the rate coefficient behavior over the full temperature range for DR onto level~4 (the third excited state) of the Se$^{2+}$ target.  That fit is not valid below 80~K, and is only accurate to within 11\% near 2\,000 and 40\,000~K (but to better than 2\% at other temperatures).  DR onto level~2 of Se$^{+}$ is also fit less accurately, to within 6\%.

\begin{figure}
  \resizebox{\hsize}{!}{\includegraphics{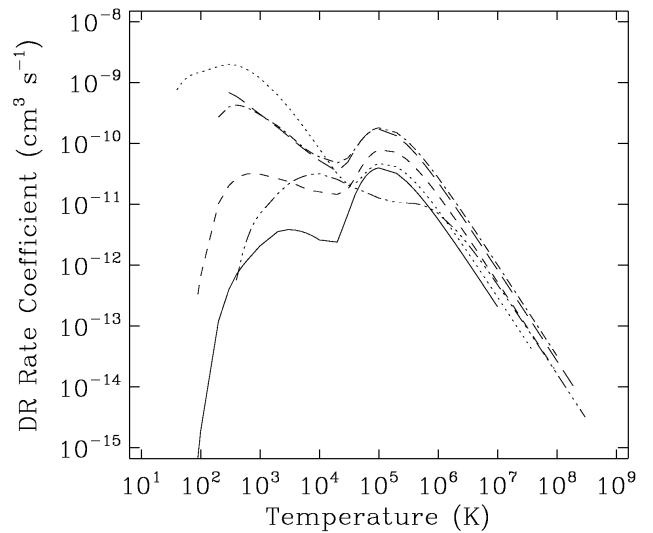}}
  \caption{Dielectronic recombination rate coefficients as a function of temperature for ground state target ions Se$^+$ (solid curve), Se$^{2+}$ (dotted curve), Se$^{3+}$ (short-dashed curve), Se$^{4+}$ (dash-dot-dash curve), Se$^{5+}$ (long-dashed curve), and Se$^{6+}$ (dash-dot-dot-dot-dash curve).} \label{drfig}
\end{figure}

\begin{table*}
\centering
\caption{Fit coefficients for dielectronic recombination rate coefficients; see Equation~\ref{dreq}.  Coefficients $c_{\rm i}$ are in cm$^3$\,s$^{-1}$\,K$^{3/2}$, and $E_{\rm i}$ are in K.  The fits are accurate to within 5\% over the temperature range (10$^1-10^7$)$z^2$~K, unless otherwise noted.  The notation $x(y)$ denotes $x\times 10^y$.}\label{drfits} 
\begin{tabular}{lcccccccc}
\hline \hline
Target & &  &  &  &  &  &  &  \\
Ion\tablefootmark{a} & Level & $c_1$ & $c_2$ & $c_3$ & $c_4$ & $c_5$ & $c_6$ & $c_7$ \\
\hline
Se$^+$ & 1 & 5.895(-8) & 1.516(-7) & 2.154(-6) & 4.509(-6) & 7.933(-4) & 2.944(-3) & 2.944(-3) \\
 & 2\tablefootmark{b} & 4.849(-9) & 4.750(-8) & 1.915(-7) & 1.839(-6) & 1.633(-4) & 2.211(-3) & 7.745(-6) \\
 & 3 & 2.580(-8) & 4.667(-8) & 2.042(-6) & 2.148(-3) & 5.616(-4) & ... & ... \\
 & 4 & 2.086(-8) & 2.435(-7) & 2.120(-6) & 3.332(-7) & 3.652(-6) & 1.713(-3) & 1.929(-4) \\
 & 5 & 2.547(-7) & 1.297(-7) & 4.131(-4) & 2.075(-3) & 2.433(-6) & ... & ... \\
Se$^{2+}$ & 1 & 4.841(-6) & 4.993(-5) & 3.566(-5) & 9.558(-5) & 9.458(-3) & 1.988(-5) & ... \\
 & 2 & 1.830(-6) & 5.312(-6) & 1.432(-5) & 2.280(-5) & 5.604(-3) & 2.176(-3) & ... \\
 & 3 & 1.686(-8) & 2.146(-7) & 5.827(-7) & 2.436(-6) & 1.391(-4) & 6.526(-3) & ... \\
 & 4\tablefootmark{c} & 3.875(-8) & 1.641(-6) & 3.522(-3) & 3.732(-3) & -3.333(-4) & ... & ... \\
 & 5 & 7.246(-13) & 9.784(-9) & 3.270(-6) & 2.470(-5) & 2.923(-4) & 1.598(-3) & 1.609(-3) \\
Se$^{3+}$ & 1 & 1.268(-6) & 2.651(-6) & 1.386(-5) & 1.166(-4) & 3.731(-3) & 6.768(-3) & 6.499(-3) \\
 & 2 & 2.286(-6) & 3.420(-6) & 4.948(-5) & 4.206(-4) & 1.030(-2) & ... & ... \\
Se$^{4+}$ & 1 & 1.505(-5) & 3.440(-5) & 2.571(-4) & 2.806(-3) & 2.952(-2) & ... & ... \\
Se$^{5+}$ & 1 & 1.107(-5) & 2.896(-5) & 8.937(-5) & 6.877(-5) & 1.457(-2) & 1.213(-2) & ... \\
Se$^{6+}$ & 1 & 2.771(-6) & 3.733(-5) & 1.459(-4) & 7.999(-4) & 5.106(-3) & 1.052(-2) & ... \\
\hline
 & & $E_1$ & $E_2$ & $E_3$ & $E_4$ & $E_5$ & $E_6$ & $E_7$ \\
\hline
Se$^+$ & 1 & 1.036(3) & 1.963(3) & 4.529(3) & 1.464(4) & 1.217(5) & 1.754(5) & 1.772(5) \\
 & 2\tablefootmark{b} & 8.714(1) & 2.241(2) & 7.538(2) & 1.046(4) & 9.325(4) & 1.525(5) & 1.764(6) \\
 & 3 & 9.716(1) & 1.663(3) & 1.077(4) & 1.330(5) & 2.180(5) & ... & ... \\
 & 4 & 2.655(2) & 3.997(2) & 9.086(2) & 1.797(3) & 3.584(4) & 1.336(5) & 2.199(5) \\
 & 5 & 8.821(2) & 2.098(3) & 9.386(4) & 1.423(5) & 8.491(5) & ... & ... \\
Se$^{2+}$ & 1 & 1.289(2) & 5.862(2) & 1.935(3) & 7.117(4) & 1.957(5) & 1.719(6) & ... \\
 & 2 & 6.188(1) & 3.923(2) & 2.220(3) & 3.978(4) & 1.797(5) & 2.263(5) & ... \\
 & 3 & 4.308(2) & 9.211(2) & 3.360(3) & 1.096(4) & 8.181(4) & 1.899(5) & ... \\
 & 4\tablefootmark{c} & 4.521(3) & 1.678(4) & 1.429(5) & 2.330(5) & 5.738(5) & ... & ... \\
 & 5 & 1.158(3) & 5.031(3) & 1.522(4) & 3.687(4) & 9.171(4) & 1.416(5) & 1.657(5) \\
Se$^{3+}$ & 1 & 7.565(2) & 1.959(3) & 6.654(3) & 3.054(4) & 1.518(5) & 2.134(5) & 2.156(5) \\
 & 2 & 6.470(2) & 3.122(3) & 1.519(4) & 6.767(4) & 1.908(5) & ... & ... \\
Se$^{4+}$ & 1 & 5.954(2) & 3.483(3) & 2.477(4) & 1.053(5) & 1.886(5) & ... & ... \\
Se$^{5+}$ & 1 & 3.417(2) & 2.862(3) & 1.556(4) & 4.233(4) & 1.420(5) & 1.941(5) & ... \\
Se$^{6+}$ & 1 & 2.567(3) & 8.368(3) & 2.392(4) & 1.260(5) & 5.466(5) & 1.186(6) & ... \\
\hline
\end{tabular}
\tablefoot{
\tablefoottext{a}{Note that the ion and level numbers correspond to the target ion (i.e., before recombination).}
\tablefoottext{b}{The fit is accurate to within 6\%.}
\tablefoottext{c}{The rate coefficient for DR onto level 4 of the Se$^{2+}$ target could not be fit to accuracies as high as other levels and ions, and is not valid below 80~K.  The fit is only accurate to within 11\% near 2\,000 and 40\,000~K, but to within 2\% at all other temperatures.}
}
\end{table*}
Fig.~\ref{drfig} illustrates the DR rate coefficients over the temperature range (10$^1-10^7$)$z^2$~K.  This figure reveals that while the rate coefficients exhibit similar behavior at high temperatures ($>10^6$~K), there are marked differences in their magnitudes and behavior at lower temperatures due to the differing resonance structure near their ionization thresholds.  

\begin{figure}
  \resizebox{\hsize}{!}{\includegraphics{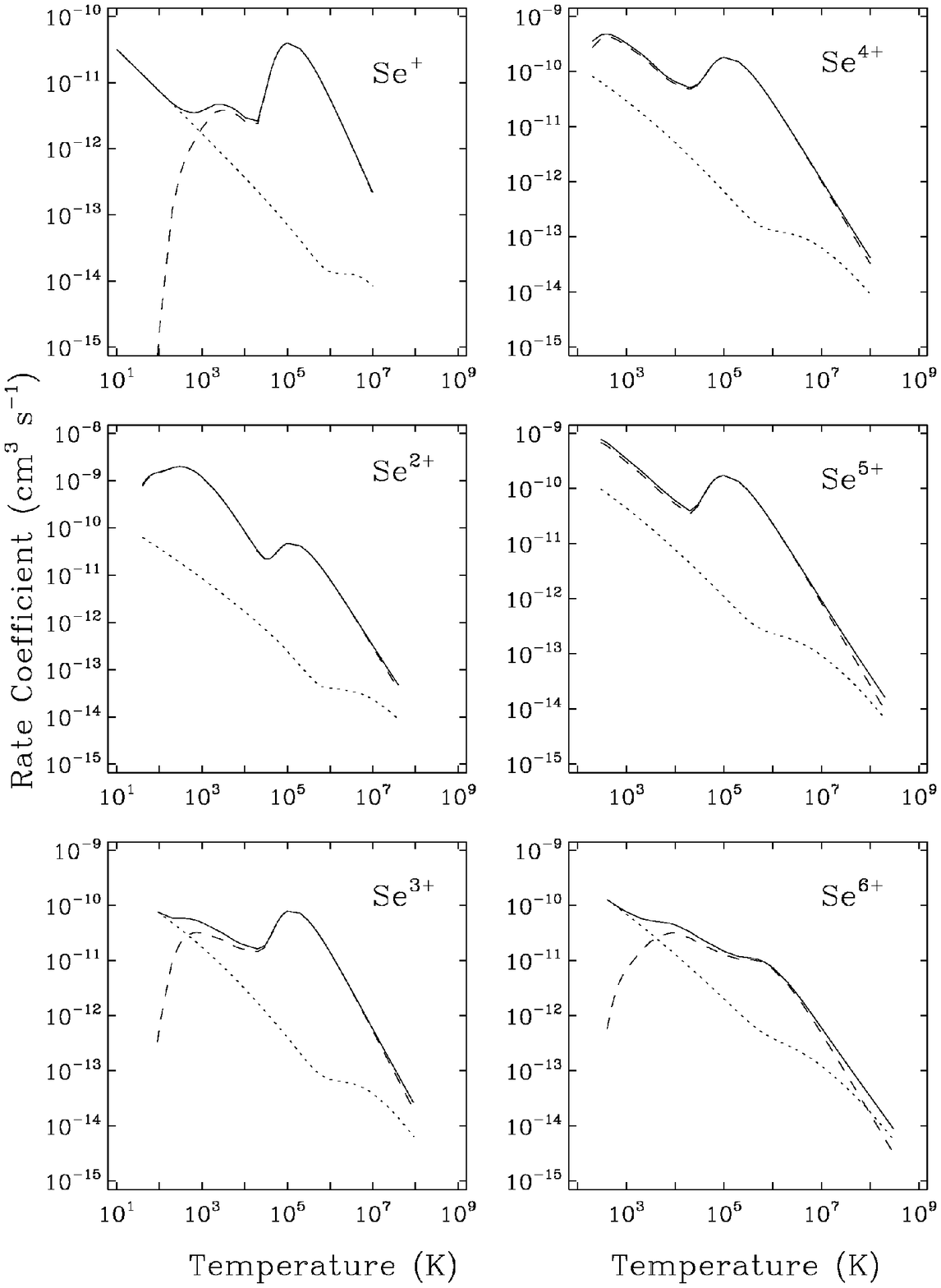}}
  \caption{Comparison of the radiative (dotted curves), dielectronic (dashed curves), and total recombination rate coefficients (solid curves) for each of the first six Se ions.} \label{totfig}
\end{figure}

Fig.~\ref{totfig} compares the RR, DR, and total recombination rate coefficients for each of the first six Se ions.  From this plot, it is apparent that DR dominates RR for each ion at 10$^4$~K, with rate coefficients larger by as much as two orders of magnitude (and occasionally even more at other temperatures).  The RR rate coefficient becomes comparable to that of DR at temperatures higher than $10^7$~K for some ions, though this is likely due to the fact that we did not consider DR from $\Delta n>0$ core excitations, which become important at high temperatures.  RR dominates DR only at low temperatures (less than 1\,000--5\,000~K) for Se$^+$, Se$^{3+}$, and Se$^{6+}$.  This highlights the importance of obtaining accurate DR rate coefficients, as it is the principal recombination process for low-charge Se ions \citep[and indeed for many other heavy elements, as noted for example by][]{badnell06a, altun07, nikolic10} in photoionized nebulae.

\subsubsection{Estimation of uncertainties}

At photoionized plasma temperatures, DR proceeds via electron capture into low-lying autoionizing states.  Unfortunately, the energies of these states have not been determined for the vast majority of atomic ions beyond the second row of the Periodic Table.  Inaccuracies in the calculated energies of these near-threshold autoionizing resonances are thus the chief source of uncertainty in computed low-temperature DR rate coefficients \citep[see, e.g., ][]{savin99, ferland03}.  While \citet{robicheaux10} showed that thermal broadening of the continuum threshold in finite density plasmas alleviates to some extent the problem of identifying which resonances lie just above the threshold (and can contribute to DR) and which do not, this effect is not likely to be sufficiently large to mitigate the comparatively large uncertainties in resonance positions resulting from theoretical calculations.

To estimate the uncertainties in our calculated DR rate coefficients due to the unknown energies of near-threshold autoionizing resonances, we shifted the continuum threshold by the amount of the largest discrepancy in calculated versus experimental target energies, with the direction of the shift determined by whether the computed energies were too large or too small relative to experiment.  For DR onto the ground state of target ions, this resulted in discrepancies in the rate coefficient near 10$^4$~K of 15--30\% for Se$^{3+}$ and Se$^{4+}$, a factor of two for Se$^{5+}$, a factor of 4--5 for Se$^{6+}$, and 1--2 orders of magnitude for Se$^{2+}$ and Se$^+$ (respectively).

We also compared the DR rate coefficients computed with the small and medium configuration sets (due to the computationally intensive nature of the DR calculations, we did not attempt to compute rate coefficients for the large configuration sets).  These uncertainties were much smaller than obtained from shifting the continuum threshold, leading to differences of 5--30\% near 10$^4$K with the exception of the Se$^+$ target (factor of 4 difference).

Clearly, the uncertainties in the DR rate coefficients are much larger than those for PI or RR.  Because DR dominates RR at photoionized plasma temperatures for these Se ions (and indeed many other heavy atomic ions), this remains an important problem for modeling photoionized astrophysical nebulae.  Since the energies of near-threshold autoionizing resonances have not been experimentally determined and are extremely difficult to accurately calculate even with the most sophisticated theoretical methods, the solution to this important problem will be challenging to achieve.  

Perhaps the most reliable method of obtaining highly accurate DR rate coefficients is electron-ion beam merging techniques with heavy-ion storage rings such as the TSR \citep{savin99, savin06, schippers10}.  Unfortunately, it is extremely challenging to use storage ring experiments to measure DR rate coefficients of low charge-to-mass species such as the Se ions investigated in this paper.  The lowest charge-to-mass ion whose recombination rate coefficient has been experimentally determined is Sc$^{3+}$, with $q/m=1/15$ \citep{schippers02}.  The maximum bending power of the dipole magnets on current storage ring facilities is the limiting factor for low $q/m$ species, which cannot be stored at sufficiently high energies.  If the ion energy is too low, then electron capture during collisions with residual gas molecules produces a very high background that prevents rate coefficient measurements \citep[S.\ Schippers, private communication; ][]{schippers02}.  Thus, experimental DR measurements of low-charge Se ions are extremely difficult if not impossible with existing storage ring facilities.  However, the forthcoming Cryogenic Storage Ring (CSR) facility in Heidelberg will use electrostatic deflectors in place of bending magnets to store the ions, and the residual gas pressure will be much smaller than in the TSR \citep{wolf06}.  The CSR will enable low-temperature DR rate coefficients to be measured for many low $q/m$ species, including those of trans-iron elements.

\section{Concluding remarks}\label{summ}

We have presented MCBP calculations of the electronic structure, distorted-wave PI cross sections (including both direct and resonant contributions), and RR and DR rate coefficients for the first six ions of the trans-iron element Se.  All of the PI, RR, and DR data are available at the CDS, and analytical fits to the RR and DR rate coefficients are presented in Tables~\ref{rrlowtfits}, \ref{rrhitfits}, and \ref{drfits}.

These atomic data are critical for efforts to determine the abundance of Se in photoionized astrophysical nebulae.  Se has been detected in dozens of planetary nebulae \citep{sterling08} and other astrophysical nebulae (see Sect.~\ref{intro}) , and is a potentially important tracer of \emph{n}-capture nucleosynthesis and the chemical evolution of trans-iron elements.

In our calculations, we tested the sensitivity of our results to the use of different configuration expansions and to other internal parameters of the AUTOSTRUCTURE code, in order to provide estimates of the (internal) uncertainties of the presented atomic data.  In the case of PI, we also compared our cross sections to recent experimental measurements \citep{esteves09, esteves10, sterling11} to gauge the accuracy of our results.  We find that the direct PI cross sections for these Se ions are uncertain by 30--50\%, while the RR rate coefficients are uncertain by $<$10\% except for Se$^+$ ($\sim$30--40\%).  Our DR rate coefficients exhibit larger uncertainties (from 15--30\% up to 1--2 orders of magnitude) due to the unknown energies of low-lying autoionizing states.  The importance of uncertainties in DR rate coefficients is highlighted by the dominance of DR over RR near 10$^4$~K, the typical temperature of photoionized nebulae such as PNe.

These uncertainties are significant, particularly for the near-neutral cases.  Low-charge Se ions are complex systems for which no comprehensive photoionization or recombination data existed prior to our study.  Thus one of our goals in addition to producing these data was to provide realistic internal uncertainties, whose effect on nebular abundance determinations can be quantified through numerical simulations of ionized nebulae.  Such an investigation will help to target the ionic systems and atomic processes that require further analysis.

This paper is the first in a series to present atomic data for the photoionization and recombination properties of trans-iron elements.  In future papers, we will present similar data for Kr and Xe, as well as charge exchange rate coefficients for low-charge ions of several \emph{n}-capture elements.  The ultimate goal of this study is to produce atomic data for \emph{n}-capture elements that are suitable to be incorporated into photoionization codes that numerically simulate the ionization and thermal structure of astrophysical nebulae.  Photoionization codes can be used to derive analytical ionization corrections needed to accurately estimate the abundances of unobserved Se, Kr, and Xe ions in ionized nebulae -- and hence to determine their elemental abundances.  Our efforts will thus enable much more accurate and robust abundance determinations of trans-iron elements in astrophysical nebulae than was previously possible.  

\acknowledgements

N.\ C.\ Sterling acknowledges support from an NSF Astronomy and Astrophysics Postdoctoral Fellowship under award AST-0901432 and from NASA grant 06-APRA206-0049.  We are grateful to N.\ R.\ Badnell for several helpful discussions regarding the use of AUTOSTRUCTURE, and for his careful reading of this manuscript.

\bibliographystyle{aa}

\bibliography{16718ms.bib}

\begin{thebibliography}{64}
\expandafter\ifx\csname natexlab\endcsname\relax\def\natexlab#1{#1}\fi

\bibitem[{Altun {et~al.}(2007)Altun, Yumak, Yavuz, {et~al.}}]{altun07}
Altun, Z., Yumak, A., Yavuz, I., {et~al.} 2007, {\aap}, 474, 1051

\bibitem[{Aspin(1994)}]{aspin94}
Aspin, C. 1994, {\aap}, 281, L29

\bibitem[{Badnell(1986)}]{badnell86}
Badnell, N.~R. 1986, {J.\ Phys.\ B: At.\ Mol.\ Opt.\ Phys.}, 19, 3827

\bibitem[{Badnell(1997)}]{badnell97}
Badnell, N.~R. 1997, {J.\ Phys.\ B: At.\ Mol.\ Opt.\ Phys.}, 30, 1

\bibitem[{Badnell(2006{\natexlab{a}})}]{badnell06a}
Badnell, N.~R. 2006{\natexlab{a}}, {\apj}, 651, L73

\bibitem[{Badnell(2006{\natexlab{b}})}]{badnell06b}
Badnell, N.~R. 2006{\natexlab{b}}, {\apjs}, 406, 1151

\bibitem[{Badnell {et~al.}(2003)Badnell, O'Mullane, Summers,
  {et~al.}}]{badnell03}
Badnell, N.~R., O'Mullane, M.~G., Summers, H.~P., {et~al.} 2003, {\aap}, 167,
  334

\bibitem[{Bahr {et~al.}(1982)Bahr, Pinnington, Kernahan, \& O'Neill}]{bahr82}
Bahr, J.~L., Pinnington, E.~H., Kernahan, J.~A., \& O'Neill, J.~A. 1982, {Can.\
  J.\ Phys.}, 60, 1108

\bibitem[{Baldwin {et~al.}(2000)Baldwin, Verner, Verner, {et~al.}}]{baldwin00}
Baldwin, J.~A., Verner, E.~M., Verner, D.~A., {et~al.} 2000, {\apjs}, 129, 229

\bibitem[{Bautista \& Kallman(2001)}]{bautista01}
Bautista, M.~A. \& Kallman, T.~R. 2001, {\apjs}, 133, 221

\bibitem[{Bi\'{e}mont \& Hansen(1986{\natexlab{a}})}]{biemont86b}
Bi\'{e}mont, E. \& Hansen, J.~E. 1986{\natexlab{a}}, {\physscr}, 34, 116

\bibitem[{Bi\'{e}mont \& Hansen(1986{\natexlab{b}})}]{biemont86a}
Bi\'{e}mont, E. \& Hansen, J.~E. 1986{\natexlab{b}}, {\physscr}, 33, 117

\bibitem[{Bi\'{e}mont \& Hansen(1987)}]{biemont87}
Bi\'{e}mont, E. \& Hansen, J.~E. 1987, {Nuc.\ Inst.\ Meth.\ Phys.\ Res.}, 23,
  274

\bibitem[{Blum \& McGregor(2008)}]{blum08}
Blum, R.~D. \& McGregor, P.~J. 2008, {\aj}, 135, 1708

\bibitem[{Burgess(1964)}]{burgess64}
Burgess, A. 1964, {\apj}, 139, 776

\bibitem[{{Busso} {et~al.}(1999){Busso}, {Gallino}, \& {Wasserburg}}]{busso99}
{Busso}, M., {Gallino}, R., \& {Wasserburg}, G.~J. 1999, {\araa}, 37, 239

\bibitem[{Chen \& Robicheaux(1994)}]{chen94}
Chen, C.-T. \& Robicheaux, F. 1994, {\pra}, 50, 3968

\bibitem[{Cowan \& Griffin(1976)}]{cowan76}
Cowan, R.~D. \& Griffin, D.~C. 1976, {J.\ Opt.\ Soc.\ Am.}, 66, 1010

\bibitem[{Curtis \& Theodosiou(1989)}]{curtis89}
Curtis, L.~J. \& Theodosiou, C.~E. 1989, {\pra}, 39, 605

\bibitem[{Esteves {et~al.}(2011)Esteves, Aguilar, Kilcoyne,
  {et~al.}}]{esteves11b}
Esteves, D.~A., Aguilar, A., Kilcoyne, A. L.~D., {et~al.} 2011, {in
  preparation}

\bibitem[{Esteves {et~al.}(2009)Esteves, Sterling, Kilcoyne,
  {et~al.}}]{esteves09}
Esteves, D.~A., Sterling, N.~C., Kilcoyne, A. L.~D., {et~al.} 2009, in {XXIX
  International Conference on Phenomena in Ionized Gases}, ed. {A.\
  Juarez-Reyes}, PA1--7, {Available at
  http://www.icpig2009.unam.mx/pdf/PA1-7.pdf}

\bibitem[{Esteves {et~al.}(2010)Esteves, Sterling, Kilcoyne,
  {et~al.}}]{esteves10}
Esteves, D.~A., Sterling, N.~C., Kilcoyne, A. L.~D., {et~al.} 2010, in {41st
  Annual Meeting of the APS Division of Atomic, Molecular and Optical Physics},
  Vol.~55, \#OPT.10

\bibitem[{Ferland(2003)}]{ferland03}
Ferland, G.~J. 2003, {\araa}, 41, 517

\bibitem[{Ferland {et~al.}(1998)Ferland, Korista, Verner, {et~al.}}]{ferland98}
Ferland, G.~J., Korista, K.~T., Verner, D.~A., {et~al.} 1998, {\pasp}, 110, 761

\bibitem[{Gibson {et~al.}(1986)Gibson, Greene, Ru\u{s}\u{c}i\'c, \&
  Berkowitz}]{gibson86}
Gibson, S.~T., Greene, J.~P., Ru\u{s}\u{c}i\'c, B., \& Berkowitz, J. 1986, {J.\
  Phys.\ B: At.\ Mol.\ Opt.\ Phys.}, 19, 2841

\bibitem[{Karakas {et~al.}(2009)Karakas, van Raai, Lugaro, Sterling, \&
  Dinerstein}]{karakas09}
Karakas, A.~I., van Raai, M.~A., Lugaro, M., Sterling, N.~C., \& Dinerstein,
  H.~L. 2009, {\apj}, 690, 1130

\bibitem[{Luhman {et~al.}(1998)Luhman, Engelbracht, \& Luhman}]{luhman98}
Luhman, K.~L., Engelbracht, C.~W., \& Luhman, M.~L. 1998, {\apj}, 499, 799

\bibitem[{Lumsden \& Puxley(1996)}]{lumsden96}
Lumsden, S.~L. \& Puxley, P.~J. 1996, {\mnras}, 281, 493

\bibitem[{Lyon {et~al.}(1986)Lyon, Peart, West, \& Dolder}]{lyon86}
Lyon, I.~C., Peart, B., West, J.~B., \& Dolder, K. 1986, {J.\ Phys.\ B: At.\
  Mol.\ Opt.\ Phys.}, 19, 4137

\bibitem[{Manson {et~al.}(1979)Manson, Msezane, Starace, \& Shahabi}]{manson79}
Manson, S.~T., Msezane, A., Starace, A.~F., \& Shahabi, S. 1979, {\pra}, 20,
  1005

\bibitem[{Marigo(2002)}]{marigo02}
Marigo, P. 2002, {\aap}, 387, 507

\bibitem[{Migdalek \& Stanek(1989)}]{migdalek89}
Migdalek, J. \& Stanek, M. 1989, {\jqsrt}, 42, 585

\bibitem[{Morton(2000)}]{morton00}
Morton, D.~C. 2000, {\apjs}, 130, 403

\bibitem[{Napiwotzki(1999)}]{napiwot99}
Napiwotzki, R. 1999, {\aap}, 350, 101

\bibitem[{Nikoli\'{c} {et~al.}(2010)Nikoli\'{c}, Gorczyca, Korista, \&
  Badnell}]{nikolic10}
Nikoli\'{c}, D., Gorczyca, T.~W., Korista, K.~T., \& Badnell, N.~R. 2010,
  {\aap}, 516, 97

\bibitem[{Okumura {et~al.}(2001)Okumura, Mori, Watanabe, Nishihara, \&
  Yamashita}]{okumura01}
Okumura, S.-I., Mori, A., Watanabe, E., Nishihara, E., \& Yamashita, T. 2001,
  {\aj}, 121, 2089

\bibitem[{Pindzola {et~al.}(1992)Pindzola, Badnell, \& Griffin}]{pindzola92}
Pindzola, M.~S., Badnell, N.~R., \& Griffin, D.~C. 1992, {\pra}, 46, 5725

\bibitem[{Puxley {et~al.}(2000)Puxley, Ramsay~Howat, \& Mountain}]{puxley00}
Puxley, P.~J., Ramsay~Howat, S.~K., \& Mountain, C.~M. 2000, {\apj}, 529, 224

\bibitem[{Ralchenko {et~al.}(2008)Ralchenko, Kramide, Reader, \& Team}]{NIST}
Ralchenko, Y., Kramide, A.~E., Reader, J., \& Team, N.~A. 2008, {NIST Atomic
  Spectra Database (version 3.1.4)}, {National Institute of Standards and
  Technology, Gaithersburg, MD.},
  {http://physics.nist.gov/cgi-bin/AtData/main\_asd3}

\bibitem[{Robicheaux {et~al.}(2010)Robicheaux, Loch, Pindzola, \&
  Ballance}]{robicheaux10}
Robicheaux, F., Loch, S.~D., Pindzola, M.~S., \& Ballance, C.~P. 2010, {\prl},
  105, 233201

\bibitem[{Roederer {et~al.}(2010)Roederer, Cowan, Karakas,
  {et~al.}}]{roederer10}
Roederer, I.~U., Cowan, J.~J., Karakas, A.~I., {et~al.} 2010, {\apj}, 724, 975

\bibitem[{Roman-Lopes {et~al.}(2009)Roman-Lopes, Abraham, Ortiz, \&
  Rodriguez-Ardila}]{roman-lopes09}
Roman-Lopes, A., Abraham, Z., Ortiz, R., \& Rodriguez-Ardila, A. 2009,
  {\mnras}, 394, 467

\bibitem[{Savin {et~al.}(2006)Savin, Gwinner, Grieser, {et~al.}}]{savin06}
Savin, D.~W., Gwinner, G., Grieser, M., {et~al.} 2006, {\apj}, 642, 1275

\bibitem[{Savin {et~al.}(1999)Savin, Kahn, Linkemann, {et~al.}}]{savin99}
Savin, D.~W., Kahn, S.~M., Linkemann, J., {et~al.} 1999, {\apjs}, 123, 687

\bibitem[{Schippers {et~al.}(2002)Schippers, Kieslich, M\"{u}ller,
  {et~al.}}]{schippers02}
Schippers, S., Kieslich, S., M\"{u}ller, A., {et~al.} 2002, {\pra}, 65, 042723

\bibitem[{Schippers {et~al.}(2011)Schippers, Lestinsky, M\"{u}ller, Savin, \&
  Wolf}]{schippers10}
Schippers, S., Lestinsky, M., M\"{u}ller, A., Savin, D.~W., \& Wolf, A. 2011,
  {Int.\ Rev.\ At.\ Mol.\ Phys.}, {in press (arXiv:1002.3678)}

\bibitem[{Sharpee {et~al.}(2007)Sharpee, Zhang, Williams, {et~al.}}]{sharpee07}
Sharpee, B., Zhang, Y., Williams, R., {et~al.} 2007, {\apj}, 659, 1265

\bibitem[{Smith \& Lambert(1990)}]{smith90}
Smith, V.~V. \& Lambert, D.~L. 1990, {\apjs}, 72, 387

\bibitem[{Sneden {et~al.}(2008)Sneden, Cowan, \& Gallino}]{sneden08}
Sneden, C., Cowan, J.~J., \& Gallino, R. 2008, {\araa}, 46, 241

\bibitem[{Stanghellini {et~al.}(2002)Stanghellini, Villaver, Manchado, \&
  Guerrero}]{stang02}
Stanghellini, L., Villaver, E., Manchado, A., \& Guerrero, M.~A. 2002, {\apj},
  576, 285

\bibitem[{Sterling \& Dinerstein(2008)}]{sterling08}
Sterling, N.~C. \& Dinerstein, H.~L. 2008, {\apjs}, 174, 158

\bibitem[{Sterling {et~al.}(2009)Sterling, Dinerstein, Hwang,
  {et~al.}}]{sterling09}
Sterling, N.~C., Dinerstein, H.~L., Hwang, S., {et~al.} 2009, {\pasa}, 26, 339

\bibitem[{Sterling {et~al.}(2007)Sterling, Dinerstein, \& Kallman}]{sterling07}
Sterling, N.~C., Dinerstein, H.~L., \& Kallman, T.~R. 2007, {\apjs}, 169, 37

\bibitem[{Sterling {et~al.}(2011)Sterling, Esteves, Bilodeau,
  {et~al.}}]{sterling11}
Sterling, N.~C., Esteves, D.~A., Bilodeau, R.~C., {et~al.} 2011, {J.\ Phys.\ B:
  At.\ Mol.\ Opt.\ Phys.}, 44, 025701

\bibitem[{Summers(2005)}]{summers05}
Summers, H.~P. 2005, {ADAS Manual v2.7}, {http://adas.phys.strath.ac.uk}

\bibitem[{Thompson {et~al.}(1978)Thompson, Lebofsky, \& Rieke}]{thompson78}
Thompson, R.~I., Lebofsky, M.~J., \& Rieke, G.~H. 1978, {\apj}, 222, L49

\bibitem[{Travaglio {et~al.}(2004)Travaglio, Gallino, Arnone,
  {et~al.}}]{travaglio04}
Travaglio, C., Gallino, R., Arnone, E., {et~al.} 2004, {\apj}, 601, 864

\bibitem[{Vanzi {et~al.}(2008)Vanzi, Cresci, Telles, \& Melnick}]{vanzi08}
Vanzi, L., Cresci, G., Telles, E., \& Melnick, J. 2008, {\aap}, 486, 393

\bibitem[{Verner \& Ferland(1996)}]{verner96a}
Verner, D.~A. \& Ferland, G.~J. 1996, {\apjs}, 103, 467

\bibitem[{Verner {et~al.}(1996)Verner, Ferland, Korista, \&
  Yakovlev}]{verner96b}
Verner, D.~A., Ferland, G.~J., Korista, K.~T., \& Yakovlev, D.~G. 1996, {\apj},
  465, 487

\bibitem[{Wallerstein {et~al.}(1997)Wallerstein, Iben, Parker,
  {et~al.}}]{wally97}
Wallerstein, G., Iben, I., Parker, P., {et~al.} 1997, {Rev.\ Mod.\ Phys.}, 69,
  995

\bibitem[{Wallerstein {et~al.}(1995)Wallerstein, Vanture, Jenkins, \&
  Fuller}]{wally95}
Wallerstein, G., Vanture, A.~D., Jenkins, E.~B., \& Fuller, G.~M. 1995, {\apj},
  449, 688

\bibitem[{Wolf {et~al.}(2006)Wolf, Buhr, Grieser, {et~al.}}]{wolf06}
Wolf, A., Buhr, H., Grieser, M., {et~al.} 2006, {Hyperfine Interactions}, 172,
  111

\bibitem[{Zatasrinny {et~al.}(2003)Zatasrinny, Gorczyca, Korista, Badnell, \&
  Savin}]{zatsarinny03}
Zatasrinny, O., Gorczyca, T.~W., Korista, K.~T., Badnell, N.~R., \& Savin,
  D.~W. 2003, {\aap}, 412, 587

\end{thebibliography}

\clearpage \onecolumn

\onltab{1}{
\begin{table}
\centering
\caption{Configuration-interaction expansions used for each Se ion.}\label{ciexp} 
\begin{tabular}{lll}
\hline \hline
Ion & Config.\ Set\tablefootmark{a} & Configurations \\
\hline
Se$^0$ & Small (10) & 4s$^2$\,4p$^4$,~~4s\,4p$^5$,~~4p$^6$ \\
 & Medium (326) & 4s$^2$\,4p$^4$,~~4s$^2$\,4p$^3$\,4d,~~4s$^2$\,4p$^3$\,4f,~~4s$^2$\,4p$^3$\,5s,~~4s$^2$\,4p$^3$\,5p, \\
 & & 4s$^2$\,4p$^2$\,4d\,5p,~~4s\,4p$^5$,~~4s\,4p$^4$\,5p,~~4s\,4p$^3$\,5p$^2$, ~~4p$^6$ \\
 & Large (785) & 4s$^2$\,4p$^4$,~~4s$^2$\,4p$^3$\,4d,~~4s$^2$\,4p$^3$\,4f,~~4s$^2$\,4p$^3$\,5s,~~4s$^2$\,4p$^3$\,5p, \\
 & & 4s$^2$\,4p$^3$\,5d,~~4s$^2$\,4p$^2$\,4d$^2$,~~4s$^2$\,4p$^2$\,4d\,5p,~~4s\,4p$^5$,~~4s\,4p$^4$\,4d, \\
 & & 4s\,4p$^4$\,5s,~~4s\,4p$^4$\,5p,~~4s\,4p$^3$\,4d\,5s,~~4s\,4p$^3$\,5p$^2$,~~ 4p$^6$, \\
 & & 3d$^9$\,4s$^2$\,4p$^5$ \\
\hline
Se$^+$ & Small (15) & 4s$^2$\,4p$^3$,~~4s\,4p$^4$,~~4p$^5$ \\
 & Medium (189) & 4s$^2$\,4p$^3$,~~4s$^2$\,4p$^2$\,4d,~~4s$^2$\,4p$^2$\,5s,~~4s$^2$\,4p$^2$\,5p,~~4s$^2$\,4p\,4d$^2$, \\
 & & 4s\,4p$^4$,~~4s\,4p$^3$\,4d,~~4p$^5$ \\
 & Large (886) & 4s$^2$\,4p$^3$,~~4s$^2$\,4p$^2$\,4d,~~4s$^2$\,4p$^2$\,5s,~~4s$^2$\,4p$^2$\,5p,~~4s$^2$\,4p$^2$\,5d, \\
 & & 4s$^2$\,4p\,4d$^2$,~~4s$^2$\,4p\,4d\,5s,~~4s$^2$\,4p\,4d\,5d,~~4s\,4p$^4$,~~4s\,4p$^3$\,4d, \\
 & & 4s\,4p$^2$\,4d$^2$,~~4p$^5$,~~4p$^4$\,4d,~~4p$^4$\,5s \\
\hline
Se$^{2+}$ & Small (32) & 4s$^2$\,4p$^2$,~~4s$^2$\,4p\,4d,~~4s\,4p$^3$,~~4p$^4$ \\
 & Medium (135) & 4s$^2$\,4p$^2$,~~4s$^2$\,4p\,4d,~~4s\,4p$^3$,~~4s\,4p$^2$\,4d,~~4p$^4$, \\
 & & 4p$^2$\,4d$^2$,~~3d$^9$\,4s$^2$\,4p$^3$ \\
 & Large (765) & 4s$^2$\,4p$^2$,~~4s$^2$\,4p\,4d,~~4s$^2$\,4p\,5s,~~4s$^2$\,4p\,5p,~~4s$^2$\,4p\,5d, \\
 & & 4s$^2$\,4d$^2$,~~4s$^2$\,4d\,5d,~~4s\,4p$^3$,~~4s\,4p$^2$\,4d,~~4s\,4p$^2$\,4f, \\
 & & 4s\,4p\,4d$^2$,~~4p$^4$,~~4p$^3$\,4d,~~4p$^3$\,5s,~~3d$^9$\,4s$^2$\,4p$^2$\,4d, \\
 & & 3d$^9$\,4s$^2$\,4p\,4d\,5s \\
\hline
Se$^{3+}$ & Small (40) & 4s$^2$\,4p,~~4s$^2$\,4d,~~4s\,4p$^2$,~~4s\,4p\,4d,~~4p$^3$ \\
 & Medium (252) & 4s$^2$\,4p,~~4s$^2$\,4d,~~4s\,4p$^2$,~~4s\,4p\,4d,~~4s\,4d$^2$, \\
 & & 4p$^3$,~~4p$^2$\,4d,~~3d$^9$\,4s$^2$\,4p\,4d,~~3d$^9$\,4s\,4p$^3$ \\
 & Large (546) & 4s$^2$\,4p,~~4s$^2$\,4d,~~4s$^2$\,5s,~~4s\,4p$^2$,~~4s\,4p\,4d, \\
 & & 4s\,4p\,4f,~~4s\,4p\,5s,~~4s\,4p\,5p,~~4s\,4p\,5d,~~4s\,4d$^2$, \\
 & & 4s\,4d\,5s,~~4s\,4d\,5p,~~4p$^3$,~~4p$^2$\,4d,~~4p$^2$\,5p, \\
 & & 4p\,4d$^2$,~~4p\,4d\,5s,~~3d$^9$\,4s$^2$\,4p\,4d,~~3d$^9$\,4s$^2$\,4d$^2$,~~3d$^9$\,4s$^2$\,4d\,5s, \\
 & & 3d$^9$\,4s\,4p$^3$ \\
\hline
Se$^{4+}$ & Small (10) & 4s$^2$,~~4s\,4p,~~4p$^2$ \\
 & Medium (47) & 4s$^2$,~~4s\,4p,~~4s\,4d,~~4s\,5s,~~4p$^2$, \\
 & & 4p\,4d,~~4d\,5d,~~5s$^2$ \\
 & Large (891) & 4s$^2$,~~4s\,4p,~~4s\,4d,~~4s\,5s,~~4s\,5p, \\
 & & 4s\,5d,~~4p$^2$,~~4p\,4d,~~4p\,5s,~~4p\,5p, \\
 & & 4d$^2$,~~4d\,5s,~~4d\,5d,~~5s$^2$,~~5p$^2$, \\
 & & 3d$^9$\,4s$^2$\,4d,~~3d$^9$\,4s$^2$\,5d,~~3d$^9$\,4s\,4p$^2$,~~3d$^9$\,4s\,4p\,4d,~~3d$^9$\,4s\,4p\,4f, \\
 & & 3d$^9$\,4s\,4f\,5p,~~3d$^9$\,4s\,5s\,5d \\
\hline
Se$^{5+}$ & Small (3) & 4s,~~4p \\
 & Medium (5) & 4s,~~4p,~~4d \\
 & Large (745) & 4s,~~4p,~~4d,~~5s,~~5p, \\
 & & 3d$^9$\,4s\,4d,~~3d$^9$\,4s\,5d,~~3d$^9$\,4p\,4d,~~3d$^9$\,4p\,4f,~~3d$^9$\,4p\,5d, \\
 & & 3d$^9$\,4d$^2$,~~3d$^9$\,4d\,5s,~~3d$^9$\,4d\,5p,~~3d$^9$\,4d\,5d,~~3d$^9$\,5s\,5p \\
\hline
Se$^{6+}$ & Small (35)\tablefootmark{b} & 3d$^{10}$,~~3d$^9$\,4s,~~3d$^9$\,4p,~~3d$^9$\,4d \\
 & Medium (35)\tablefootmark{b} & 3d$^{10}$,~~3d$^9$\,4s,~~3d$^9$\,4p,~~3d$^9$\,4d \\
 & Large (57) & 3d$^{10}$,~~3d$^9$\,4s,~~3d$^9$\,4p,~~3d$^9$\,4d,3d$^9$\,5s, \\
 & & 3d$^9$\,5d \\
\hline
\end{tabular}
\tablefoot{
\tablefoottext{a}{After the name of each configuration set, the number of levels is listed in parentheses.}
\tablefoottext{b}{Although the same CI expansion was utilized for the small and medium sets of Se$^{6+}$, the scaling parameters of the medium set were optimized differently: the configurations 3d$^9$\,5d, 3d$^9$\,6d, and 3d$^9$\,7d were included in the configuration list to determine scaling parameters, but removed in the final structure calculation.}
}
\end{table}
}

\onllongtab{2}{
\centering
\begin{longtable}{lccc}
\caption{Radial scaling parameters used for CI expansions of each ion}\label{lambdas} \\
\hline \hline
\multicolumn{4}{c}{Se$^0$} \\
\hline
Orbital & Small & Medium & Large \\
\hline
\endfirsthead
\caption{Continued.} \\
\hline
\endhead
\endfoot
\hline
\endlastfoot
1s & 1.25936 & 1.25385 & 1.25535 \\
2s & 1.09800 & 1.09813 & 1.09987 \\
2p & 1.06516 & 1.06543 & 1.06445 \\
3s & 1.02964 & 1.02983 & 1.03142 \\
3p & 1.01691 & 1.01707 & 1.01706 \\
3d & 0.98901 & 0.98953 & 0.98912 \\
4s & 0.96050 & 0.96355 & 0.96485 \\
4p & 0.97914 & 0.98363 & 0.98332 \\
4d & ... & 1.00469 & 1.03918 \\
4f & ... & 1.02392 & 1.01451 \\
5s & ... & 0.97193 & 1.03068 \\
5p & ... & 1.03442 & 1.02918 \\
5d & ... & ... & 1.01438 \\
\hline
\multicolumn{4}{c}{Se$^+$} \\
\hline
Orbital & Small & Medium & Large \\
\hline
1s & 1.25216 & 1.25516 & 1.25364 \\
2s & 1.09851 & 1.09631 & 1.09585 \\
2p & 1.06482 & 1.06447 & 1.06307 \\
3s & 1.03101 & 1.02970 & 1.03059 \\
3p & 1.01607 & 1.01610 & 1.01536 \\
3d & 0.98874 & 0.98929 & 0.98970 \\
4s & 0.96953 & 0.96867 & 0.97226 \\
4p & 0.97901 & 0.98050 & 0.97058 \\
4d & ... & 0.99909 & 0.99017 \\
5s & ... & 0.98123 & 0.98732 \\
5p & ... & 1.01508 & 1.02418 \\
5d & ... & ... & 1.03819 \\
\hline
\multicolumn{4}{c}{Se$^{2+}$} \\
\hline
Orbital & Small & Medium & Large \\
\hline
1s & 1.26607 & 1.25544 & 1.25389 \\
2s & 1.09894 & 1.09742 & 1.09713 \\
2p & 1.06384 & 1.06381 & 1.06429 \\
3s & 1.03177 & 1.03118 & 1.03111 \\
3p & 1.01549 & 1.01509 & 1.01696 \\
3d & 0.98797 & 0.98820 & 0.98923 \\
4s & 0.97518 & 0.98274 & 0.97642 \\
4p & 0.98166 & 0.98982 & 0.99900 \\
4d & 0.98458 & 0.98379 & 1.00493 \\
4f & ... & ... & 1.09951 \\
5s & ... & ... & 0.98270 \\
5p & ... & ... & 1.01451 \\
5d & ... & ... & 1.02193 \\
\hline
\multicolumn{4}{c}{Se$^{3+}$} \\
\hline
Orbital & Small & Medium & Large \\
\hline
1s & 1.25706 & 1.25957 & 1.25060 \\
2s & 1.09599 & 1.09701 & 1.09890 \\
2p & 1.06234 & 1.06303 & 1.06278 \\
3s & 1.02897 & 1.03152 & 1.03118 \\
3p & 1.01422 & 1.01569 & 1.01614 \\
3d & 0.98716 & 0.98855 & 0.98899 \\
4s & 0.97684 & 0.97863 & 0.98035 \\
4p & 0.98730 & 1.00378 & 1.00134 \\
4d & 0.97966 & 1.00165 & 1.00353 \\
4f & ... & ... & 1.04938 \\
5s & ... & ... & 1.01012 \\
5p & ... & ... & 0.97718 \\
5d & ... & ... & 0.97798 \\
\hline
\multicolumn{4}{c}{Se$^{4+}$} \\
\hline
Orbital & Small & Medium & Large \\
\hline
1s & 1.25396 & 1.26031 & 1.26001 \\
2s & 1.09493 & 1.09398 & 1.09585 \\
2p & 1.06204 & 1.06105 & 1.06283 \\
3s & 1.02798 & 1.02729 & 1.02992 \\
3p & 1.01276 & 1.01272 & 1.01453 \\
3d & 0.98639 & 0.98662 & 0.98858 \\
4s & 0.97686 & 0.97774 & 0.98177 \\
4p & 0.97413 & 0.98345 & 0.97896 \\
4d & ... & 0.96499 & 0.99410 \\
4f & ... & ... & 1.03288 \\
5s & ... & 1.00779 & 1.02644 \\
5p & ... & ... & 0.96971 \\
5d & ... & ... & 0.99887 \\
\hline
\multicolumn{4}{c}{Se$^{5+}$} \\
\hline
Orbital & Small & Medium & Large \\
\hline
1s & 1.26114 & 1.25370 & 1.25921 \\
2s & 1.09560 & 1.09502 & 1.09944 \\
2p & 1.06089 & 1.05993 & 1.06193 \\
3s & 1.02749 & 1.02797 & 1.03420 \\
3p & 1.01136 & 1.01121 & 1.01392 \\
3d & 0.98493 & 0.98512 & 0.98800 \\
4s & 0.97805 & 0.97821 & 0.98381 \\
4p & 0.97409 & 0.97389 & 1.00989 \\
4d & ... & 0.96765 & 1.00662 \\
4f & ... & ... & 1.01505 \\
5s & ... & ... & 1.01864 \\
5p & ... & ... & 1.01194 \\
5d & ... & ... & 0.99808 \\
\hline
\multicolumn{4}{c}{Se$^{6+}$} \\
\hline
Orbital & Small & Medium & Large \\
\hline
1s & 1.25708 & 1.25307 & 1.25533 \\
2s & 1.09531 & 1.09680 & 1.09484 \\
2p & 1.05923 & 1.05934 & 1.05967 \\
3s & 1.02647 & 1.02599 & 1.02605 \\
3p & 1.00974 & 1.00941 & 1.00948 \\
3d & 0.98305 & 0.98296 & 0.98285 \\
4s & 1.00424 & 1.00419 & 1.00427 \\
4p & 0.99619 & 0.99617 & 0.99622 \\
4d & 0.97412 & 0.97553 & 0.97555 \\
5s & ... & ... & 0.99360 \\
5d & ... & ... & 0.97139 \\
\end{longtable}
}

\renewcommand{\thefootnote}{\alph{footnote}}

\onllongtab{3}{
\centering
\begin{longtable}{ccccccc}
\caption{Comparison of selected calculated and experimental energies (in Rydbergs)}\label{ecomp} \\
\hline \hline
\multicolumn{7}{c}{Se$^0$} \\
\hline
Index & Config. & Term & Small & Medium & Large & NIST \\
\hline
\endfirsthead
\caption{Continued.} \\
\endhead
\endfoot
\hline
\endlastfoot
1 & 4s$^2$\,4p$^4$ & $^3$P$_2$ & 0.0000 & 0.0000 & 0.0000 & 0.0000 \\
2 &  & $^3$P$_1$ & 0.0172 & 0.0177 & 0.0177 & 0.0181 \\
3 &  & $^3$P$_0$ & 0.0223 & 0.0225 & 0.0227 & 0.0231 \\
4 &  & $^1$D$_2$ & 0.1085 & 0.1034 & 0.1053 & 0.0873 \\
5 &  & $^1$S$_0$ & 0.1960 & 0.1826 & 0.1929 & 0.2045 \\
6 & 4s$^2$\,4p$^3$\,5s & $^5$S$_2$ & ... & 0.3945 & 0.4302 & 0.4391 \\
7 &  & $^3$S$_1$ & ... & 0.4307 & 0.4791 & 0.4647 \\
8 & 4s$^2$\,4p$^3$\,5p & $^5$P$_1$ & ... & 0.4964 & 0.5225 & 0.5399 \\
9 &  & $^5$P$_2$ & ... & 0.4974 & 0.5233 & 0.5403 \\
10 &  & $^5$P$_3$ & ... & 0.4991 & 0.5249 & 0.5412 \\
11 & 4s$^2$\,4p$^3$\,4d & $^5$D$_2$ & ... & 0.5347 & 0.5533 & 0.5776 \\
12 &  & $^5$D$_1$ & ... & 0.5347 & 0.5533 & 0.5775 \\
13 &  & $^5$D$_3$ & ... & 0.5347 & 0.5533 & 0.5776 \\
14 &  & $^5$D$_0$ & ... & 0.5347 & 0.5532 & 0.5776 \\
15 &  & $^5$D$_4$ & ... & 0.5348 & 0.5534 & 0.5776 \\
16 & 4s$^2$\,4p$^3$\,5p & $^3$P$_1$ & ... & 0.5396 & 0.5636 & 0.5524 \\
17 &  & $^3$P$_2$ & ... & 0.5404 & 0.5645 & 0.5529 \\
18 &  & $^3$P$_0$ & ... & 0.5409 & 0.5649 & 0.5804 \\
19 & 4s$^2$\,4p$^3$\,4d & $^3$D$_1$ & ... & 0.5452 & 0.5760 & 0.5951 \\
20 &  & $^3$D$_2$ & ... & 0.5462 & 0.5764 & 0.5949 \\
21 &  & $^3$D$_3$ & ... & 0.5478 & 0.5776 & 0.5954 \\
22 & 4s$^2$\,4p$^3$\,5s & $^3$D$_2$ & ... & 0.5569 & 0.5907 & 0.5634 \\
23 &  & $^3$D$_1$ & ... & 0.5571 & 0.5902 & 0.5621 \\
24 &  & $^3$D$_3$ & ... & 0.5584 & 0.5925 & 0.5672 \\
25 &  & $^1$D$_2$ & ... & 0.5701 & 0.6134 & 0.5785 \\
I.P. & ... & ... & 0.6377 & 0.6558 & 0.6590 & 0.7168 \\
\hline
\multicolumn{7}{c}{Se$^+$} \\
\hline
Index & Config. & Term & Small & Medium & Large & NIST \\
\hline
1 & 4s$^2$\,4p$^3$ & $^4$S$_{3/2}$ & 0.0000 & 0.0000 & 0.0000 & 0.0000 \\
2 &  & $^2$D$_{3/2}$ & 0.1527 & 0.1470 & 0.1413 & 0.1200 \\
3 &  & $^2$D$_{5/2}$ & 0.1581 & 0.1520 & 0.1457 & 0.1256 \\
4 &  & $^2$P$_{1/2}$ & 0.2204 & 0.2218 & 0.2104 & 0.2099 \\
5 &  & $^2$P$_{3/2}$ & 0.2267 & 0.2277 & 0.2156 & 0.2177 \\
6 & 4s\,4p$^4$ & $^4$P$_{5/2}$ & 0.7851 & 0.7569 & 0.7460 & 0.7643 \\
7 &  & $^4$P$_{3/2}$ & 0.8001 & 0.7708 & 0.7582 & 0.7799 \\
8 &  & $^4$P$_{1/2}$ & 0.8081 & 0.7783 & 0.7647 & 0.7877 \\
9 & 4s$^2$\,4p$^2$\,5s & $^4$P$_{1/2}$ & ... & 0.9045 & 0.8912 & 0.8682 \\
10 &  & $^4$P$_{3/2}$ & ... & 0.9145 & 0.9002 & 0.8817 \\
11 &  & $^4$P$_{5/2}$ & ... & 0.9289 & 0.9129 & 0.8992 \\
12 &  & $^2$P$_{1/2}$ & ... & 0.9433 & 0.9254 & 0.9012 \\
13 &  & $^2$P$_{3/2}$ & ... & 0.9523 & 0.9276 & 0.9236 \\
14 &  & $^2$D$_{5/2}$ & ... & 1.0396 & 1.0192 & 0.9874 \\
15 &  & $^2$D$_{3/2}$ & ... & 1.0399 & 1.0187 & 0.9883 \\
16 & 4s$^2$\,4p$^2$\,5p & $^2$S$_{1/2}$ & ... & 1.0541 & 1.0644 & 1.0699 \\
I.P. & ... & ... & 1.5128 & 1.5242 & 1.5388 & 1.5574 \\
\hline
\multicolumn{7}{c}{Se$^{2+}$} \\
\hline
Index & Config. & Term & Small & Medium & Large & NIST \\
\hline
1 & 4s$^2$\,4p$^2$ & $^3$P$_0$ & 0.0000 & 0.0000 & 0.0000 & 0.0000 \\
2 &  & $^3$P$_1$ & 0.0135 & 0.0142 & 0.0158 & 0.0000 \\
3 &  & $^3$P$_2$ & 0.0315 & 0.0334 & 0.0362 & 0.0000 \\
4 &  & $^1$D$_2$ & 0.1383 & 0.1406 & 0.1407 & 0.1188 \\
5 &  & $^1$S$_0$ & 0.2424 & 0.2710 & 0.2595 & 0.2591 \\
6 & 4s\,4p$^3$ & $^3$D$_1$ & 0.7935 & 0.8376 & 0.8308 & 0.8301 \\
7 &  & $^3$D$_2$ & 0.7937 & 0.8378 & 0.8313 & 0.8450 \\
8 &  & $^3$D$_3$ & 0.7967 & 0.8412 & 0.8353 & 0.8798 \\
9 &  & $^3$P$_2$ & 0.9267 & 0.9727 & 0.9654 & 0.9706 \\
10 &  & $^3$P$_0$ & 0.9274 & 0.9735 & 0.9675 & 0.9703 \\
11 &  & $^3$P$_1$ & 0.9275 & 0.9737 & 0.9676 & 0.9713 \\
12 & 4s$^2$\,4p\,4d & $^1$D$_2$\footnotemark[1] & 1.0250 & 1.0685 & 1.0262 & 1.2685 \\
13 &  & $^3$F$_2$ & 1.1291 & 1.1691 & 1.1314 & 1.1304 \\
14 &  & $^3$F$_3$ & 1.1390 & 1.1795 & 1.1424 & 1.1419 \\
15 &  & $^3$F$_4$ & 1.1543 & 1.1962 & 1.1602 & 1.1610 \\
16 &  & $^1$P$_1$ & 1.2586 & 1.3019 & 1.3215 & 1.2479 \\
17 &  & $^3$P$_2$ & 1.2984 & 1.3365 & 1.3042 & 1.3004 \\
18 &  & $^3$P$_1$ & 1.3067 & 1.3453 & 1.3310 & 1.3009 \\
19 &  & $^3$P$_0$ & 1.3156 & 1.3559 & 1.3283 & 1.2941 \\
20 &  & $^3$D$_1$ & 1.3292 & 1.3673 & 1.3054 & 1.2816 \\
21 &  & $^3$D$_3$ & 1.3331 & 1.3701 & 1.3255 & 1.2969 \\
22 &  & $^3$D$_2$ & 1.3338 & 1.3718 & 1.3328 & 1.2704 \\
23 &  & $^1$F$_3$ & 1.4080 & 1.4427 & 1.3855 & 1.3548 \\
24 & 4s\,4p$^3$ & $^1$D$_2$\footnotemark[1] & 1.4366 & 1.4801 & 1.4289 & 1.0258 \\
I.P. & ... & ... & 2.2322 & 2.2591 & 2.2585 & 2.2653 \\
\hline
\multicolumn{7}{c}{Se$^{3+}$} \\
\hline
Index & Config. & Term & Small & Medium & Large & NIST \\
\hline
1 & 4s$^2$\,4p & $^2$P$_{1/2}$ & 0.0000 & 0.0000 & 0.0000 & 0.0000 \\
2 &  & $^2$P$_{3/2}$ & 0.0359 & 0.0408 & 0.0399 & 0.0399 \\
3 & 4s\,4p$^2$ & $^4$P$_{1/2}$ & 0.6690 & 0.6954 & 0.7242 & 0.7235 \\
4 &  & $^4$P$_{3/2}$ & 0.6826 & 0.7110 & 0.7394 & 0.7379 \\
5 &  & $^4$P$_{5/2}$ & 0.7025 & 0.7332 & 0.7612 & 0.7617 \\
6 &  & $^2$D$_{3/2}$ & 0.9631 & 0.9645 & 0.9848 & 0.9496 \\
7 &  & $^2$D$_{5/2}$ & 0.9649 & 0.9675 & 0.9878 & 0.9542 \\
8 &  & $^2$S$_{1/2}$ & 1.2240 & 1.2380 & 1.2570 & 1.1736 \\
9 &  & $^2$P$_{1/2}$ & 1.3662 & 1.3312 & 1.3517 & 1.2405 \\
10 &  & $^2$P$_{3/2}$ & 1.3883 & 1.3537 & 1.3735 & 1.2608 \\
11 & 4s$^2$\,4d & $^2$D$_{3/2}$ & 1.4846 & 1.4520 & 1.4404 & 1.3962 \\
12 &  & $^2$D$_{5/2}$ & 1.4868 & 1.4542 & 1.4425 & 1.3998 \\
13 & 4p$^3$ & $^2$D$_{3/2}$ & 1.7948 & 1.8299 & 1.8413 & 1.7288 \\
14 &  & $^2$D$_{5/2}$ & 1.7999 & 1.8360 & 1.8473 & 1.7948 \\
16 &  & $^2$P$_{3/2}$ & 2.0512 & 2.0962 & 2.0974 & 1.9987 \\
17 &  & $^2$P$_{1/2}$ & 2.0536 & 2.0988 & 2.0977 & 2.0644 \\
I.P. & ... & ... & 3.0904 & 3.1181 & 3.1140 & 3.1564 \\
\hline
\multicolumn{7}{c}{Se$^{4+}$} \\
\hline
Index & Config. & Term & Small & Medium & Large & NIST \\
\hline
1 & 4s$^2$ & $^1$S$_0$ & 0.0000 & 0.0000 & 0.0000 & 0.0000 \\
2 & 4s\,4p & $^3$P$_0$ & 0.7789 & 0.7712 & 0.8130 & 0.8179 \\
3 &  & $^3$P$_1$ & 0.7912 & 0.7859 & 0.8255 & 0.8325 \\
4 &  & $^3$P$_2$ & 0.8179 & 0.8188 & 0.8529 & 0.8654 \\
5 &  & $^1$P$_1$ & 1.2842 & 1.2353 & 1.2567 & 1.2004 \\
6 & 4p$^2$ & $^1$D$_2$ & 2.0516 & 1.8880 & 1.9396 & 1.9428 \\
7 &  & $^3$P$_0$ & 1.8912 & 1.9003 & 1.9485 & 1.9300 \\
8 &  & $^3$P$_1$ & 1.9072 & 1.9207 & 1.9655 & 1.9509 \\
9 &  & $^3$P$_2$ & 1.9303 & 1.9585 & 1.9973 & 1.9922 \\
10 & 4s\,4d & $^3$D$_1$ & ... & 2.2974 & 2.3557 & 2.3468 \\
11 &  & $^3$D$_2$ & ... & 2.2996 & 2.3577 & 2.3488 \\
12 &  & $^3$D$_3$ & ... & 2.3029 & 2.3608 & 2.3518 \\
I.P. & ... & ... & 4.8651 & 4.8835 & 4.9191 & 5.0211 \\
\hline
\multicolumn{7}{c}{Se$^{5+}$} \\
\hline
Index & Config. & Term & Small & Medium & Large & NIST \\
\hline
1 & 4s & $^2$S$_{1/2}$ & 0.0000 & 0.0000 & 0.0000 & 0.0000 \\
2 & 4p & $^2$P$_{1/2}$ & 1.0065 & 1.0063 & 1.0272 & 1.0276 \\
3 &  & $^2$P$_{3/2}$ & 1.0521 & 1.0520 & 1.0840 & 1.0795 \\
4 & 4d & $^2$D$_{3/2}$ & ... & 2.5111 & 2.5749 & 2.5773 \\
5 &  & $^2$D$_{5/2}$ & ... & 2.5168 & 2.5826 & 2.5835 \\
I.P. & ... & ... & 5.8878 & 5.8863 & 5.9592 & 6.0052 \\
\hline
\multicolumn{7}{c}{Se$^{6+}$} \\
\hline
Index & Config. & Term & Small & Medium & Large & NIST \\
\hline
1 & 3d$^{10}$ & $^1$S$_0$ & 0.0000 & 0.0000 & 0.0000 & 0.0000 \\
2 & 3d$^9$\,4s & $^3$D$_3$ & 4.0123 & 4.0164 & 4.0173 & 4.0073 \\
3 &  & $^3$D$_2$ & 4.0351 & 4.0393 & 4.0401 & 4.0289 \\
4 &  & $^3$D$_1$ & 4.0731 & 4.0772 & 4.0781 & 4.0712 \\
5 &  & $^1$D$_2$ & 4.1138 & 4.1179 & 4.1187 & 4.1035 \\
6 & 3d$^9$\,4p & $^3$P$_2$ & 5.0873 & 5.0914 & 5.0929 & 5.0913 \\
7 &  & $^3$F$_3$ & 5.1359 & 5.1400 & 5.1415 & 5.1360 \\
8 &  & $^3$P$_1$ & 5.1468 & 5.1509 & 5.1524 & 5.1545 \\
9 &  & $^3$F$_4$ & 5.1720 & 5.1762 & 5.1777 & 5.1771 \\
10 &  & $^3$F$_2$ & 5.1760 & 5.1801 & 5.1816 & 5.1809 \\
11 &  & $^3$P$_0$ & 5.1813 & 5.1854 & 5.1869 & 5.1922 \\
12 &  & $^1$D$_2$ & 5.2463 & 5.2505 & 5.2520 & 5.2408 \\
13 &  & $^1$F$_3$ & 5.2465 & 5.2507 & 5.2522 & 5.2448 \\
14 &  & $^3$D$_3$ & 5.2703 & 5.2745 & 5.2760 & 5.2705 \\
15 &  & $^1$P$_1$ & 5.2816 & 5.2858 & 5.2873 & 5.2696 \\
16 &  & $^3$D$_1$ & 5.3055 & 5.3098 & 5.3113 & 5.2989 \\
17 &  & $^3$D$_2$ & 5.3128 & 5.3170 & 5.3185 & 5.3104 \\
I.P.\footnotemark[2] & ... & ... & ... & ... & ... & 11.4209 \\
\footnotetext[1]{The internal identification scheme within AUTOSTRUCTURE likely mislabeled the Se$^{2+}$ 4s$^2$\,4p\,4d and 4s\,4p$^3$ $^1$D$_2$ terms, leading to the large discrepancies in calculated vs.\ experimental energies.  If the identifications are exchanged, the energy agreement improves significantly.}
\footnotetext[2]{Since we did not compute the electronic structure of Se$^{7+}$, it was not possible to calculate the ionization potential of Se$^{6+}$.}
\end{longtable}
}

\renewcommand{\thefootnote}{\alph{footnote}}
\onllongtab{4}{
\centering
\begin{longtable}{llcccc}
\caption{Calculated Einstein A-coefficients (in s$^{-1}$) compared to literature values.  A number in parentheses indicates a power of ten (i.e., 5.0(-10)~=~5.0$\times10^{-10}$).}\label{acomp} \\
\hline \hline
\multicolumn{6}{c}{Se$^0$} \\
\hline
$i$ & $j$ & Small & Medium & Large & Previous\footnotemark[1] \\
\hline
\endfirsthead
\caption{Continued.} \\
\hline
\endhead
\hline
\endfoot
\hline
\endlastfoot
1 & 2 & 1.50(-1) & 1.64(-1) & 1.64(-1) & 1.71(-1) \\
1 & 3 & 1.53(-4) & 4.53(-4) & 4.29(-4) & 1.96(-4) \\
2 & 3 & 9.28(-3) & 7.38(-3) & 8.58(-3) & 1.03(-3) \\
1 & 4 & 7.59(-1) & 7.84(-1) & 7.97(-1) & 6.40(-1) \\
2 & 4 & 1.50(-1) & 1.47(-1) & 1.51(-1) & 1.08(-1) \\
3 & 4 & 5.71(-4) & 6.25(-4) & 5.28(-4) & 7.86(-5) \\
1 & 5 & 4.50(-2) & 1.26(-2) & 2.46(-2) & 1.89(-1) \\
2 & 5 & 6.83 & 6.88 & 7.26 & 7.61 \\
4 & 5 & 5.26(-1) & 5.47(-1) & 8.44(-1) & 2.44 \\
1 & 6 & ... & 7.41(5) & 7.57(5) & 1.70(6) \\
2 & 6 & ... & 1.36(5) & 1.44(5) & 3.24(5) \\
1 & 7 & ... & 7.92(8) & 7.92(9) & 2.13(8) \\
2 & 7 & ... & 4.00(8) & 4.01(8) & 9.79(7) \\
3 & 7 & ... & 1.30(8) & 1.31(8) & 3.30(7) \\
1 & 19 & ... & 4.42(5) & 1.26(7) & 1.32(8) \\
1 & 20 & ... & 9.38(6) & 1.30(8) & 1.46(7) \\
1 & 21 & ... & 1.69(8) & 6.61(8) & 5.28(8) \\
\hline
\multicolumn{6}{c}{Se$^+$} \\
\hline
$i$ & $j$ & Small & Medium & Large & Previous\footnotemark[1] \\
\hline
1 & 2 & 3.21(-1) & 2.72(-1) & 1.97(-1) & 3.50(-1) \\
1 & 3 & 2.33(-2) & 2.32(-2) & 1.84(-2) & 1.88(-2) \\
2 & 3 & 2.13(-3) & 1.66(-3) & 1.10(-3) & 2.78(-3) \\
1 & 4 & 1.61 & 1.63 & 1.24 & 2.30 \\
2 & 4 & 5.87(-1) & 7.10(-1) & 5.31(-1) & 1.23 \\
3 & 4 & 2.30(-2) & 5.09(-2) & 4.00(-2) & 9.37(-2) \\
1 & 5 & 3.54 & 3.64 & 2.78 & 4.95 \\
2 & 5 & 1.10 & 1.24 & 9.13(-1) & 2.24 \\
3 & 5 & 5.67(-1) & 6.98(-1) & 5.22(-1) & 1.25 \\
4 & 5 & 2.67(-3) & 2.32(-3) & 1.57(-3) & 4.91(-3) \\
1 & 9 & ... & 1.37(9) & 1.52(9) & 9.96(8) \\
1 & 10 & ... & 1.41(9) & 1.56(9) & 1.03(9) \\
1 & 11 & ... & 1.50(9) & 1.65(9) & 1.14(9) \\
\hline
\multicolumn{6}{c}{Se$^{2+}$} \\
\hline
$i$ & $j$ & Small & Medium & Large & Previous\footnotemark[1] \\
\hline
1 & 2 & 5.72(-2) & 6.75(-2) & 9.19(-2) & 9.59(-2) \\
1 & 3 & 9.63(-5) & 1.15(-4) & 1.84(-4) & 1.66(-4) \\
2 & 3 & 1.03(-1) & 1.23(-1) & 1.48(-1) & 1.35(-1) \\
1 & 4 & 1.81(-3) & 1.36(-3) & 2.27(-3) & 2.37(-4) \\
2 & 4 & 6.21(-1) & 7.34(-1) & 8.70(-1) & 8.00(-1) \\
3 & 4 & 1.15 & 1.32 & 1.50 & 1.21 \\
2 & 5 & 1.22(1) & 1.57(1) & 1.69(1) & 1.70(1) \\
3 & 5 & 1.42(-1) & 3.38(-1) & 3.62(-1) & 5.81(-1) \\
4 & 5 & 7.46(-1) & 2.17 & 1.57 & 2.81 \\
\hline
\multicolumn{6}{c}{Se$^{3+}$} \\
\hline
$i$ & $j$ & Small & Medium & Large & Previous\footnotemark[1] \\
\hline
1 & 2 & 5.51(-1) & 8.06(-1) & 7.57(-1) & 7.58(-1) \\
1 & 6 & 2.78(8) & 2.39(8) & 2.66(8) & 2.79(8) \\
2 & 6 & 2.06(7) & 1.14(7) & 1.33(7) & 4.90(7) \\
2 & 7 & 2.62(8) & 2.14(8) & 2.41(8) & 2.98(8) \\
\hline
\multicolumn{6}{c}{Se$^{4+}$} \\
\hline
$i$ & $j$ & Small & Medium & Large & Previous\footnotemark[1] \\
\hline
1 & 3 & 2.30(6) & 4.14(6) & 4.13(6) & 1.24(7) \\
1 & 5 & 8.04(9) & 7.23(9) & 8.50(9) & 1.79(10) \\
\hline
\multicolumn{6}{c}{Se$^{5+}$} \\
\hline
$i$ & $j$ & Small & Medium & Large & Previous\footnotemark[1] \\
\hline
1 & 2 & 2.31(9) & 2.31(9) & 2.56(9) & 2.22(9) \\
1 & 3 & 2.64(9) & 2.64(9) & 3.01(9) & 5.07(9) \\
2 & 4 & ... & 9.67(9) & 1.05(10) & 1.95(10) \\
3 & 4 & ... & 1.76(9) & 1.88(9) & 1.80(9) \\
3 & 5 & ... & 1.07(10) & 1.14(10) & 1.64(10) \\
\footnotetext[1]{References for forbidden transitions in the ground configuration: Se$^0$ \citep{biemont86b}, Se$^+$ and Se$^{2+}$ \citep{biemont86a}, Se$^{3+}$ \citep{biemont87}.  \citet{morton00} is the reference for Se$^0$--Se$^{3+}$ permitted transitions, while those for Se$^{4+}$ and Se$^{5+}$ are taken from \citet{migdalek89}, \citet{bahr82}, and \citet{curtis89}.  To our knowledge, tabulated Se$^{6+}$ Einstein A-coefficients have not been published.}
\end{longtable}
}

\end{document}